\documentclass[sigconf,preprint]{acmart}

\usepackage{graphicx}
\usepackage{booktabs} 
\usepackage{multirow}
\usepackage{tabularx, makecell}
\usepackage{listings}
\usepackage{array}
\usepackage{xcolor,colortbl}
\usepackage{dblfloatfix}
\usepackage{caption}
\usepackage{subcaption}
\usepackage[compact]{titlesec}
\titlespacing{\section}{0pt}{5pt}{0pt}
\titlespacing{\subsection}{0pt}{5pt}{0pt}
\titlespacing{\subsubsection}{0pt}{5pt}{0pt}
\AtBeginDocument{
  \setlength\abovedisplayskip{0pt}
  \setlength\belowdisplayskip{0pt}
}
\begin{document}

\title{An Exploratory Study of Debugging Episodes
}

\author{Abdulaziz Alaboudi}
\affiliation{%
  \institution{George Mason University}
  \city{Fairfax, Virginia}
  \country{USA}}
\email{aalaboud@gmu.edu}

\author{Thomas D. LaToza}
\affiliation{%
  \institution{George Mason University}
  \city{Fairfax, Virginia}
  \country{USA}}
\email{tlatoza@gmu.edu}


\begin{abstract}
Many studies have long investigated how developers debug, shaping our understanding of debugging and helping motivate the creation of more effective tools. However, less is known about the typical progression of debugging in real world settings. In this study, we focus on characterizing debugging episodes from the moment at which developers first encounter a defect to the moment at which it is resolved. We investigate the typical duration and frequency of debugging episodes and the typical activities which occur. We observed developers by watching professional developers at work in live-streamed programming sessions. Using this data source, we curated 15 sessions in which 11 professional developers worked for 30 hours. We then systematically coded the debugging episodes and activities that occurred within these videos, yielding a dataset of 2137 debugging activities and 1407 programming activities. We found that debugging was frequent, even in programming work, occurring once every eight minutes. Debugging episodes vary greatly in time, with most being less than a few minutes and a few as more than 100 minutes. However, most debugging time is spent in long debugging episodes. We found no single activity that dominated debugging time, and long debugging episodes often involved many diverse activities. Finally, we found that,in terms of the activities developers did, programming and debugging were remarkably similar, particularly in the frequency of editing and browsing code. 


\end{abstract}
\maketitle

\section{Introduction}
For decades, debugging has been the subject of much research work. Researchers have investigated debugging in terms of what makes it hard \cite{eisenstadt1993tales, Layman-debugging-revisited, CokerQualitative2019}, the kinds of questions developers ask \cite{Ko2007InformationNeeds, LaToza2010ReachabilityQuestions}, categories of strategies developers follow \cite{BohmeFSE-DebuggingHypotheses, katz1987debugging}, and types of debugging tools developers use \cite{Murphy2006, petrillo2019swarm}. These studies have revealed many important aspects of debugging, opening new opportunities to improve the way we teach \cite{mccauley2008debugging} and design tools for debugging.

However, prior studies of debugging are limited in several important ways. 
Most studies of debugging have been conducted either in the lab or through the analysis of log data from instrumented development environments. 
Lab settings allow researchers to observe developers while debugging after receiving a defect report. However, lab settings are inherently artificial, reflecting an unfamiliar context in which developers work on unfamiliar code. While log data enables understanding debugging in naturalistic contexts, it is inherently limited in its ability to reveal developer intent. While log data can effectively answer questions about how frequently various tools are used, it is much more difficult to answer questions developers' intent when using each tool. 

In this paper, we conducted the first \textit{exploratory study} of debugging episodes in a naturalistic setting. 
A debugging episode begins when a developer first encounters a defect while programming or receives a defect report. Throughout the episode, developers perform activities, where they edit, browse, test, inspect code, and consult online resources. As they do so, they navigate between files, edit files, and use tools to inspect program state. A debugging episodes ends when the developers fixes the defect or decides to stop.

\begin{table*}[]
\centering
\caption{A summary of prior studies of developer debugging behavior.}
\label{tab:lit}
\resizebox{\textwidth}{!}{%
\begin{tabular}{lllclll}
\hline
Authors &
  Study type &
  Data &
  Participants &
  Organizations &
  Duration \{hours\} &
  Observed behavior \\ \toprule
Afzal et.al \cite{Afzal2018} &
  Field &
  Logs &
  81 &
  Unknown &
  15K &
  Debugger usage \\
Murphy et.al \cite{Murphy2006} &
  Field &
  Logs &
  41 &
  Unknown &
  Unknown &
  Debugger usage \\
Damevski et.al \cite{Damevski2017} &
  Field &
  Logs &
  196 &
  1 &
  33K &
  Debugger usage \\
Amann et.al \cite{Amann2016} &
  Field &
  Logs &
  84 &
  1 &
  6.3K &
  Debugger usage \\
Petrillo et.al \cite{petrillo2019swarm}        & Controlled & Logs, Observation & 28 & Mostly students & 10      & Debugger usage                \\
Beller et.al \cite{beller2018dichotomy} &
  Field &
  Logs, self-reports &
  634 &
  Unknown &
  18K &
  Debugger usage \\
Perscheid et.al \cite{Perscheid2017} &
  Field &
  Observation &
  8 &
  4 &
  Unkown &
  Strategies and debugger usage \\
Chattopadhyay et.al \cite{Chattopadhyay2019} &
  Field &
  Observation &
  3 &
  1 &
  2 &
  Strategies \\
  Ko et.al \cite{Ko2007InformationNeeds} &
  Field &
  Observation &
  17&
  1 &
  25 &
  Information seeking \\
LaToza et.al \cite{LaToza2010ReachabilityQuestions} &
  Field &
  Observation &
  17&
  1 &
  25 &
  Information seeking \\
  Vans et.al \cite{vans1999program} &
  Field &
  Observation &
  4&
  1 &
  8 &
  Strategies \\
Lawrance et.al \cite{Lawrance2013IFT} &
  Controlled &
  Observation &
  10 &
  1 &
  20 &
  Strategies \\
Jiang et.al \cite{jiang2017programmers} &
  Controlled &
  Observation &
  9 &
  Mostly students &
  18 &
  Strategies \\
Baltes et.al \cite{baltes2015navigate} &
  Controlled &
  Observation &
  12 &
  Mostly students &
  12 &
  Strategies \\
Gould et.al \cite{gould1975some} &
  Controlled &
  Observation &
  10 &
  1 &
  Unknown &
  Strategies \\
Vessey \cite{Vessey1985} &
  Controlled &
  Observation &
  16 &
  1 &
  6 &
  Strategies \\
Katz et.al \cite{katz1987debugging} &
  Controlled &
  Observation &
  77 &
  Mostly students &
  Unknown &
  Strategies \\
  Gugerty et.al \cite{Gugerty86} &
  Controlled &
  Observation &
  44 &
  Mostly students &
  Unknown &
  Strategies \\
Layman et.al \cite{Layman-debugging-revisited} & Field      & self-reports      & 15 & 1               & Unknown & Strategies and debugger usage \\
Böhme et.al \cite{BohmeFSE-DebuggingHypotheses} &
  Controlled &
  self-reports &
  12 &
  Unknown &
  22 &
  Strategies \\ \bottomrule
\end{tabular}%
}
\end{table*}

To understand the nature of debugging episodes, we observed developers. We observed developers debugging software projects in varying sizes, programming languages, practices, and domains. To curate such a diverse dataset, we used a recently available new source of data: live-streamed programming videos \cite{AlaboudiVLHCC2019}. In these videos, developers work and think aloud to explain how and why they are working as they do. Researchers have found that these videos show professional developers working in real-time on open source projects \cite{AlaboudiVLHCC2019, Alaboudi2019CHASE}, podcasting hours development and debugging work on projects used in production.
Using this data source, we curated 15 sessions (each corresponding to a separate video) in which 11 professional developers worked for 30 hours. Our analysis of this dataset yielded 89 distinct debugging episodes spanning more than 15 hours of debugging time as well as 13 hours of programming time. We analyzed the developers' activities during debugging episodes and programming work, resulting in 2137 and 1407 activities for debugging and programming, respectively. We focused on answering the following research questions:

\begin{itemize}
\item[\textbf{RQ1}] What is the frequency and duration of debugging episodes?
 
\item[\textbf{RQ2}] What are the characteristics of debugging activities?

\item[\textbf{RQ3}] What characteristics differentiate long from short debugging episodes? 
 
\item[\textbf{RQ4}] How does the time spent on activities vary between debugging and programming?
\end{itemize}

We found that debugging episodes varied widely in duration range, from a few seconds to more than a hundred minutes, with a skewed distribution. Most debugging time was spend in the longest 25\% of debugging episodes. Debugging occurred frequently in programming work, occurring after every eight minutes of programming. There was no single activity that consumed the majority of debugging time. Long debugging episodes involved a wide range of activities, including editing and browsing files, inspect the program state, and consulting online resources. Short debugging episodes consisted primarily of editing and testing edits. Finally, we found that programming and debugging were remarkably similar in terms of the activities developers did, particularly in editing and browsing code. The largest differences were in how developers tested and inspected code.

To support researchers in using live-streamed programming videos to study developers in real-world settings, we built the observe-dev.online platform. It includes a dataset of more than 100 hours of programming work by 33 professional developers.
We discuss our platform in section \ref{sec:Observe.dev}.

\section{Related Work}
Many studies have long investigated debugging behavior, beginning at least as early as 1974 \cite{Gould1974}(Table \ref{tab:lit}). These studies encompass both field studies examining behavior in a naturalistic context as well as experiments investigating debugging in a controlled setting. Many studies directly observing developers, while others have indirectly observed debugging through log data or self-reports made by developers. However, only five studies have directly observed developers in a field setting.
These studies have examined debugging from a wide range of perspectives, including the strategies developers use, the use of the debugger, challenges developers face, and the time developers spend debugging.

Developers use a variety of debugging strategies. Developers often follow a simple hypothesis testing strategy \cite{Perscheid2017, zayour2016qualitative}, speculating about the cause of a defect based on their experience with similar defects \cite{eisenstadt1993tales, BohmeFSE-DebuggingHypotheses} and clues they gather about the program behavior, state, or output \cite{gould1975some, Gugerty86}.
Developers then test their hypotheses by editing their code \cite{Zeller2005, Layman-debugging-revisited} or inspecting the program's behavior and state \cite{gould1975some, Perscheid2017}. 
Developers use both forwards and backwards reasoning strategies.
In forwards reasoning, developers follow the execution of the failing test \cite{BohmeFSE-DebuggingHypotheses}, building a mental representation of the program \cite{katz1987debugging} and inspecting program execution and state through breakpoints \cite{romero2007debugging}. 
In backwards reasoning \cite{gould1975some, BohmeFSE-DebuggingHypotheses,lukey1980understanding}, developers start from the incorrect output and work backwards in the execution to the defect location. Information foraging theory (IFT) models how developers navigate code \cite{Lawrance2013IFT, piorkowski2015fix, piorkowski2013whats}, including in debugging tasks. According to IFT, developers navigate between patches (e.g., methods) based on their scent (e.g., method identifiers), which offer hints directing developers to their prey (e.g., the fault location). Developers more frequently switch between subgoals when debugging than in programming~\cite{Chattopadhyay2019}.
Experienced developers are able to make use of their knowledge to comprehend code at a higher level of abstraction than novices~\cite{Vessey1985, vans1999program}. 

An important source of data for studying debugging are logs. Logs instrument programming and debugging tools to record what developers do in real-world programming work. Log data enables studying debugging at scale, enabling individual studies to examine as much as 18,000 hours of developer activity~\cite{beller2018dichotomy}. Log data has enabled extensive study of the use of debuggers in practice.
Debuggers have been found to be among the most used features in modern IDEs \cite{Murphy2006, Amann2016}, which developers use at beginning of a debugging episode \cite{Afzal2018}. Developers often avoid complex debugger features such as breakpoints \cite{Damevski2017} and prefer simpler debugging techniques such as ``printf debugging'' \cite{beller2018dichotomy}.
However, log data has important limitations \cite{myers2016programmers}. In  recording only developer actions without additional context, it can be difficult or impossible to reconstruct the developers' intent and determine, for example, if developer actions relate to a debugging or programming task.

A number of studies have enumerated specific \textit{challenges} developers face that can make debugging difficult. An analysis of debugging "war stories" found the two most common causes of difficulty were inapplicable debugging tools and ``large temporal or spatial chasms between the root cause and the symptom''~\cite{eisenstadt1993tales}. Developers face difficulties reproducing defects and determining the root cause of failures~\cite{Ko2007InformationNeeds}.
Modern systems' multithreaded and distributed nature can make instrumentation and testing debugging hypotheses challenging~\cite{Layman-debugging-revisited}. API's degree of abstraction impose unique challenges for debugging~\cite{CokerQualitative2019}.
Other studies have identified specific questions developers report to be hard-to-answer or that are associated with particularly time-consuming debugging episodes~\cite{LaToza2010ReachabilityQuestions, LaToza2010Hard-to-answerCode}.

Studies have also been conducted to quantify the time developers spend debugging, yielding widely varying results. Beller et al. \cite{beller2018dichotomy} observed that developers spent only 14\% of their active IDE time in the debugger. Minelli et.al \cite{Minelli2016} and Meyer et.al \cite{meyer2014software} also reported a low percentage (0.87\%, and 3.9\% respectively) of total time using the debugger. However, developers reported that they spent between 20\%-60\% of the working time debugging, which researchers have argued to be an over-estimation\cite{beller2018dichotomy}.

Other studies have focused on evaluating the impact of various debugging aids on the productivity of developers in debugging tasks. For example, automatic debugging tools model the debugging task as a search problem to identify a defect location, shrinking the search space developers need to inspect \cite{MarkWeiser1984ProgramSlicing, DeMillo1996, Zhang2006, XiangyuZhang2003, jones2002visualization}. Studies measuring the impact of these tools on developers' debugging performance have found mixed benefits, suggesting the need for support beyond localizing the defect \cite{Parnin2011AreAutomatedDebugging, Wang2015EvaluatingIR, Alaboudi2020}.

\begin{table*}[]
\centering
\caption{Our dataset of 15 sessions. If a session was less than two hours, we included another session from the same developer.}
\label{tab:my-table}
\begin{tabular}{cccllcc}
\multicolumn{2}{c}{Developer} & \multirow{2}{*}{\begin{tabular}[c]{@{}c@{}}\\ Session Time\end{tabular}} & \multicolumn{4}{c}{Project} \\ \cline{1-2} \cline{4-7} 
ID & Yrs. Exp.{\footnotesize{$^*$}} &         & Name and Brief Description                & LOC             & Files & Commits \\ \toprule
D1   & 10 & 2:00:56 & \textit{Firefox}: A popular web browser.     & 4M JavaScript   & 6K    & 73M     \\
D2   & 31 & 2:40:48 & \textit{Curl}: A library for transferring data.                & 138K C          & 704   & 26K     \\
D3 &
  7 &
1:23:56 + 1:22:14 &
  \textit{Serenity OS}:  A Unix-like operating system. &
 120K C++ &
  1.3K &
  15K 
  \\
D4   & 8  & 3:28:35 & \textit{Tox}:  A library for task automation.                  & 10.7K Python    & 88    & 2K      \\
D5   & 8  & 2:59:49 & \textit{Downshift}:  A set of web components built with React. & 11K JavaScript  & 82    & 652\\
D6 &
  8 &
  1:18:38 + 1:05:10  &
  \textit{Uzual}:  A mobile app that helps track daily habits &
  2.5K JavaScript &
  45 &
  204 \\

D7   & 10 & 2:54:43 & \textit{Vectrexy}: A game emulator.                            & 9.7K C++        & 90    & 656     \\
D8 &
  9 &
  1:21:06 + 1:44:19 &
  \textit{Kap}: A screen recorder for computers. &
  9.5K JavaScript &
  12 &
  866 \\
D9   & 9  & 2:15:26 & \textit{DevBette}:  A web application for a small business.    & 3.6K C\#        & 84    & 287     \\
D10 &
  10 &
  1:13:40 + 1:03:34 &
  \textit{Alacritty}:  A cross-platform terminal. &
  16.5K Rust &
  63 &
  1.7K \\
D11  & 8  & 3:40:00 & \textit{Webpack}:  A bundler for javascript.                   & 95K JavaScript  & 3.1K  & 12.6K   \\ 
        \bottomrule
      \multicolumn{6}{c}{\footnotesize{$^*$ \# of years contributing to open source projects.}}
\end{tabular}
\label{tab:dataset}
\end{table*}

This paper offers the first study of debugging episodes conducted in a naturalistic setting.
Our work offers new findings on the duration, frequency, and activities of debugging and the differences between activity in programming and debugging. 

\begin{figure}
    \centering
    \includegraphics[scale=.25, width=1\linewidth,trim=0cm 0cm 0cm 0cm,clip]{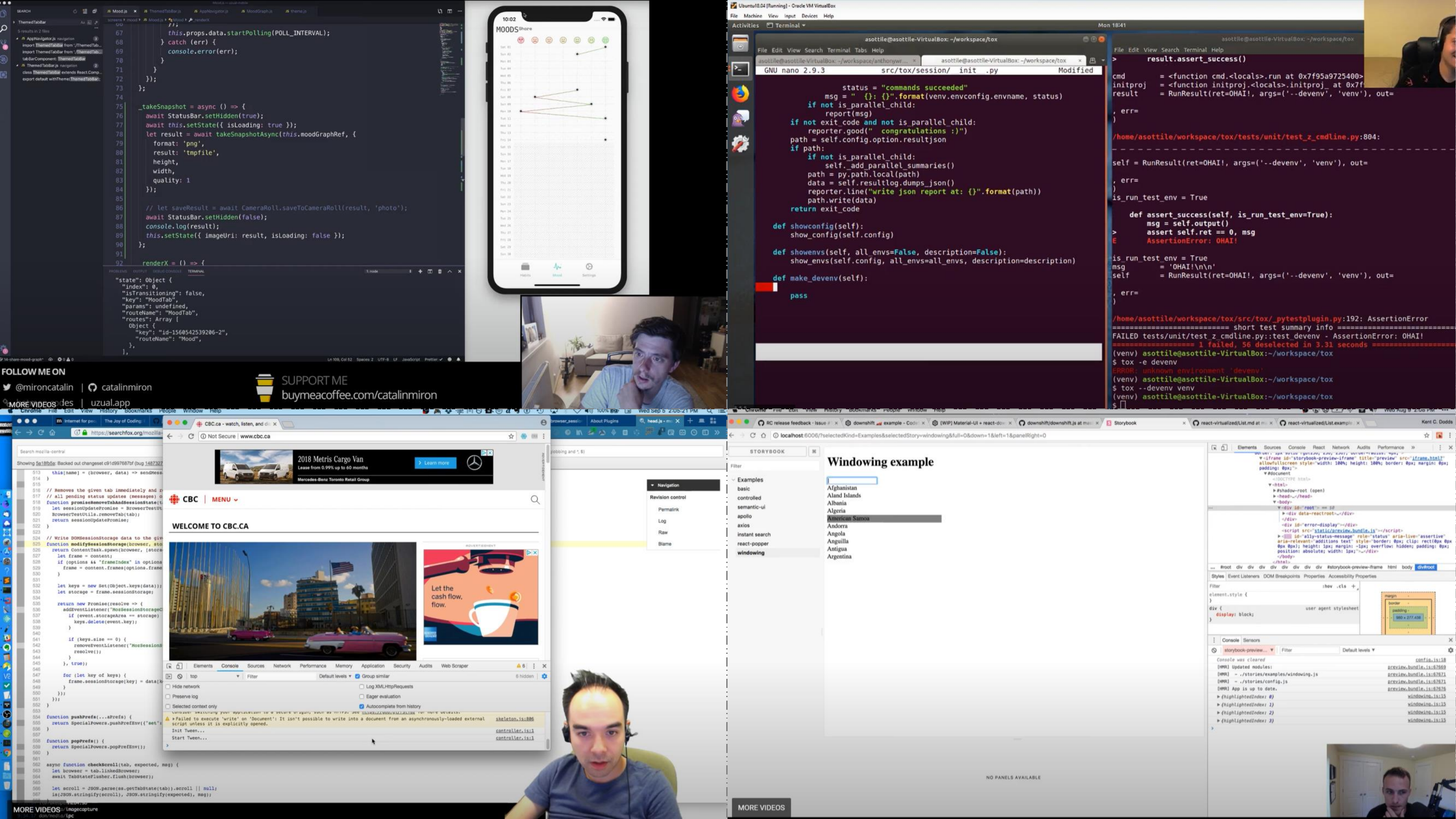}
    \caption{Developers live-stream their programming work on open source projects, which they record and share on platforms such as YouTube.}
    \label{fig:examples}
\end{figure}

\section{Method}
To investigate debugging episodes, we observed professional developers through live-streamed programming videos. For brevity, we refer to each live-streamed programming video as a \textit{session}. We collected sessions reflecting a diverse cross-section of software projects in varying sizes, programming languages, and domains. We then developed a coding scheme to identify specific episodes and activities within the sessions to explore debugging episodes. Finally, we coded the sessions using the coding scheme.

\subsection{Live-Streamed Programming Videos}
Using platforms such as YouTube and Twitch, some developers have recently begun the practice of live-streaming their work, broadcasting and recording their real-time work contributing to open source software projects \cite{Faas2018}. 
Researchers have found that these sessions are not rehearsed and illustrate developers' moment-to-moment work contributing to real software projects using their preferred development environment  \cite{AlaboudiVLHCC2019} (Figure \ref{fig:examples}). Moreover, as they explain their work to watchers, the sessions document both developers' actions and a running commentary, similar to think-aloud, describing why they are working as they do. Observational studies in software engineering have traditionally been limited by the difficulty of gaining access to software developers at work on real projects and the impossibility of sharing datasets due to confidentiality. Much as widespread access by researchers to the repositories of open source projects or questions and answer on Stack Overflow has led to the proliferation of empirical software engineering \cite{lakhani2004, Mamykina2011, singer2014Twitter, MacLeod2015, chatterjee2019exploratory}, we believe these sessions offer a similar opportunity, complimenting these datasets by offering the ability to answer new questions where direct observation is required. To enable this opportunity, we have constructed an open source platform for researchers working with live-streamed programming videos, which we describe in Section \ref{sec:Observe.dev}. 

Sessions share a common session structure \cite{AlaboudiVLHCC2019}. 
Developers start the live-stream by stating a goal for the session. 
They then work towards this goal, reading documentation and writing, debugging, and running code. 
Most videos depict work on open source projects, which developers link to in their session descriptions. 
Developers are occasionally interrupted, either by developers watching live or by others in their physical space, mirroring the typical interruptions developers face in their day-to-day work \cite{meyer2014software, abad2018task}. 
After completing the session, developers may archive the video on platforms such as YouTube and Twitch, with most licensed under the Creative Commons.

\subsection{Data Collection}
To select sessions, we formulated a strict data collection methodology. 
To find a relevant session, we used YouTube search functionality with keywords such as  ``open source contribution'' and ``live-stream programming'' as well live-streamed programming communities on Reddit\footnote{https://www.reddit.com/r/WatchPeopleCode/} and Github\footnote{https://github.com/bnb/awesome-developer-streams} for links to sessions.  
In selecting videos to include, we used three criteria:
\begin{itemize}
    \item \textit{The archived session is available}: Many developers use Twitch to host their sessions. However, Twitch only archives videos for 14 days. As we needed the ability to conduct analyses over an extended period, we excluded sessions that were not archived on a long-term basis. 
    
    \item \textit{The project is open source and ready to be used by other projects or users}: We first checked that the developer's project was hosted in a public code repository. After locating the project repository, we skimmed the documentation and issue tracker, looking for evidence that it had or will have a public release for general usage. For example, we looked for dates describing a future release or a link to an executable version of the project. 
    
    \item \textit{The video shows significant development work}: To ensure the video contained a meaningful length of development work, we briefly skimmed each session. We excluded videos where developers primarily spent the session time communicating with other developers through chat. When a session was shorter than two hours, we chose another session from the same developer working on the same project. 
\end{itemize}

We sought to create a diverse sample of sessions, encompassing developers using a variety of programming languages and working in a variety of application domains. 
Recent field studies of debugging have observed between eight and ten developers employed by one to four different companies  \cite{Perscheid2017, Chattopadhyay2019}. Based on this, we chose to 
observe eleven developers working on eleven distinct open-source projects in sessions of at least two hours. 
Our dataset includes desktop applications, command-line programs, mobile apps, web apps, games, and operating systems. Table \ref{tab:dataset} lists the 15 sessions. As a conservative estimate of developers' experience, we examined each developers' GitHub profile page and identified their first commit to an open source project.  
All eleven developers actively worked on open source projects for a period of 7 to 31\footnote{GitHub allows developers to migrate their open source contributions from other legacy version control systems, resulting in commits that predate GitHub.} years (median = 9 years). 
Some shared their current or past employer in their GitHub profiles, including Google, Microsoft, Lyft, PayPal, and Mozilla. 
The total duration of these sessions is 30 hours.

\begin{figure}
    \centering
    \includegraphics[scale=.5]{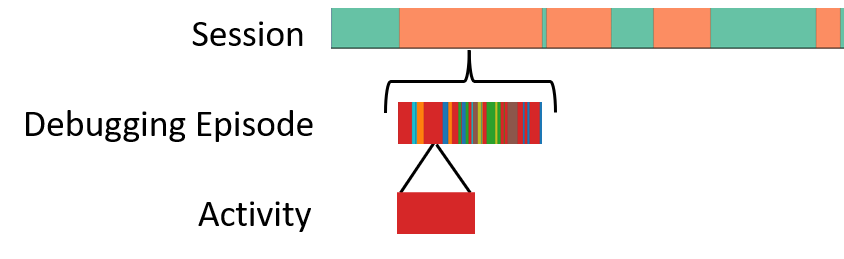}
    \caption{Our analysis focused on identifying debugging episodes and their activities within a session.}
    \label{fig:session}
\end{figure}

\begin{table*}[]
\centering
\caption{Summary of the coding book for activities. We coded the activities for debugging episodes and programming work.}
 \label{tab:codingBook}
\resizebox{\textwidth}{!}{%
\begin{tabular}{llll}
\hline
Activity               & Start Criteria                                           & End Criteria                                             & Characteristics                             \\ \toprule
Browsing A File of Code  & \multirow{2}{*}{Open a file of code.}           & Leave the file without edits.                   & Duration, file name, file type            \\
Editing A File of Code   &                                                 & Leave the file after editing.                   & Duration, file name, file type, edit type \\
Testing Program          & Run a program to observe its final output.      & Leave test environment                                 & Duration, testing method, test target       \\
Inspecting Program State & Run a program to observe program state values.  & Leave inspect env.                              & Duration, inspection method                 \\
Consulting External Resources & Open the browser or local doc to search for info. & Leave the browser or the resource. & Duration, resources type, causes \\
Other                    & Engage in noncoding work (e.g., writing notes). & Engage in noncoding work (e.g., writing notes). & Duration, work type  \\                \bottomrule   
\end{tabular}%
}
\end{table*}
\subsection{Data Analysis} 
To identify debugging episodes and their activities within a session (Figure \ref{fig:session}), we collected different definition of debugging \cite{johnson1982software,ko2011state,Parnin2011AreAutomatedDebugging} to guide our identification of a debugging episode. 
The first author watched ten sessions iterative and built an initial codebook that contained the definition of a debugging episode and six different activities. The complete codebook is available in the replication package\footnote{https://figshare.com/s/6c4026a941db49ea3de3}.

Due to a space limitation, we briefly summarize our coding book. A Debugging episode starts when one of three events happens during a session. First, An error message appears in the console,  program UI or falling tests. Second, developers express variably that the output is not what they expect. Third, developers start reproducing a defect reported in the issue the tracker. Developers have to address the defect to mark a duration within a session as the start of the debugging episode. The debugging episode ends if one of two events occurs. First, the error disappears, or the program produced the expected output. Second, developers are verbally stating that they stopped debugging. We considered the rest of the session that does not include debugging episodes as programming work. To avoid coding irrelevant work such as break or social chatting, we exclude these segments of the session and mark them as irrelevant. 
 
We coded activities for both programming work and debugging episodes. These activities were capturing the types of work a developer did during a segment of time. These included five code-focused activities as well as an ``other'' activity containing non-code-focused work, such as interacting with the development environment, checking the version control system, and writing notes. Table \ref{tab:codingBook} summarizes the activities and the characteristics we coded for each. 


After constructing the initial codebook, the two authors iteratively coded sessions, discussing disagreements after each iteration and revising the codebook. Instead of coding a single session, we choose several representative sessions with differing codebases, programming languages, and development tools. Through these iterations, the two authors coded 20 distinct episodes and 166 activities. The last iteration yielded a Chen's Kappa inter-rater agreement \cite{Richard1977} of 75\% for episodes and 84\% for activities, reflecting substantial and almost perfect agreement, respectively. 
Using the final codebook, the first author then coded the entire dataset using observe.dev-online (Section \ref{sec:Observe.dev}).
The entire dataset, including the videos and codes, is publicly available\footnote{https://bit.ly/3kkbL2W}.


\section{Results}

\begin{figure}
    \centering
    \includegraphics[scale=.28,trim= 7.5cm 3cm 6.5cm 6cm, ,clip]{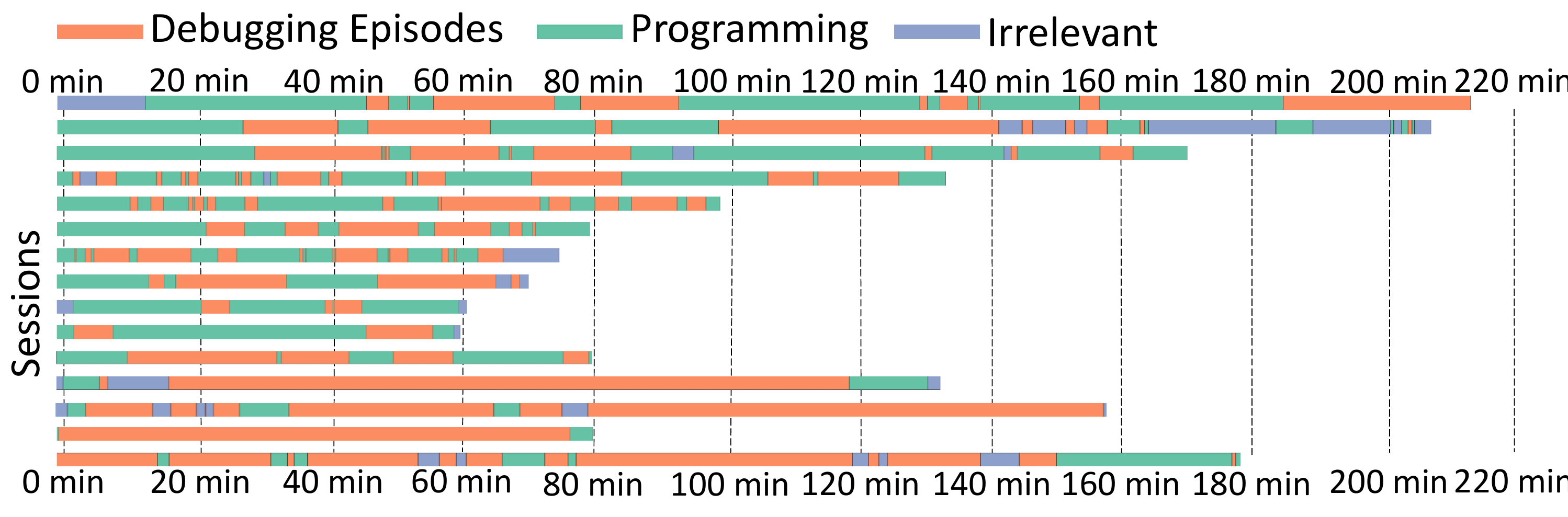}
    \caption{ Debugging episodes and programming in the 15 sessions. }
    \label{fig:debuggingAndprogramming}
\end{figure}

We coded 1407 activities in 13 hours of programming work and 2137 activities in 15 hours of debugging work, including 89 distinct debugging episodes. The average number of activities per debugging episode was 24 ($\pm$37)\footnote{$\pm$ is used to report stander deviation.}. Figure \ref{fig:debuggingAndprogramming} shows  debugging episodes and programming work in the 15 sessions. Our exploratory analysis reveled observations regarding frequency and duration of debugging episodes (RQ1), activities of debugging episodes (RQ2), differences in activities between long and short episodes (RQ3), and differences in activities between debugging episodes and programming work (RQ4).   

\begin{figure}
    \centering
    \includegraphics[scale = .12, trim= 0cm 5cm 0cm 15cm,clip]{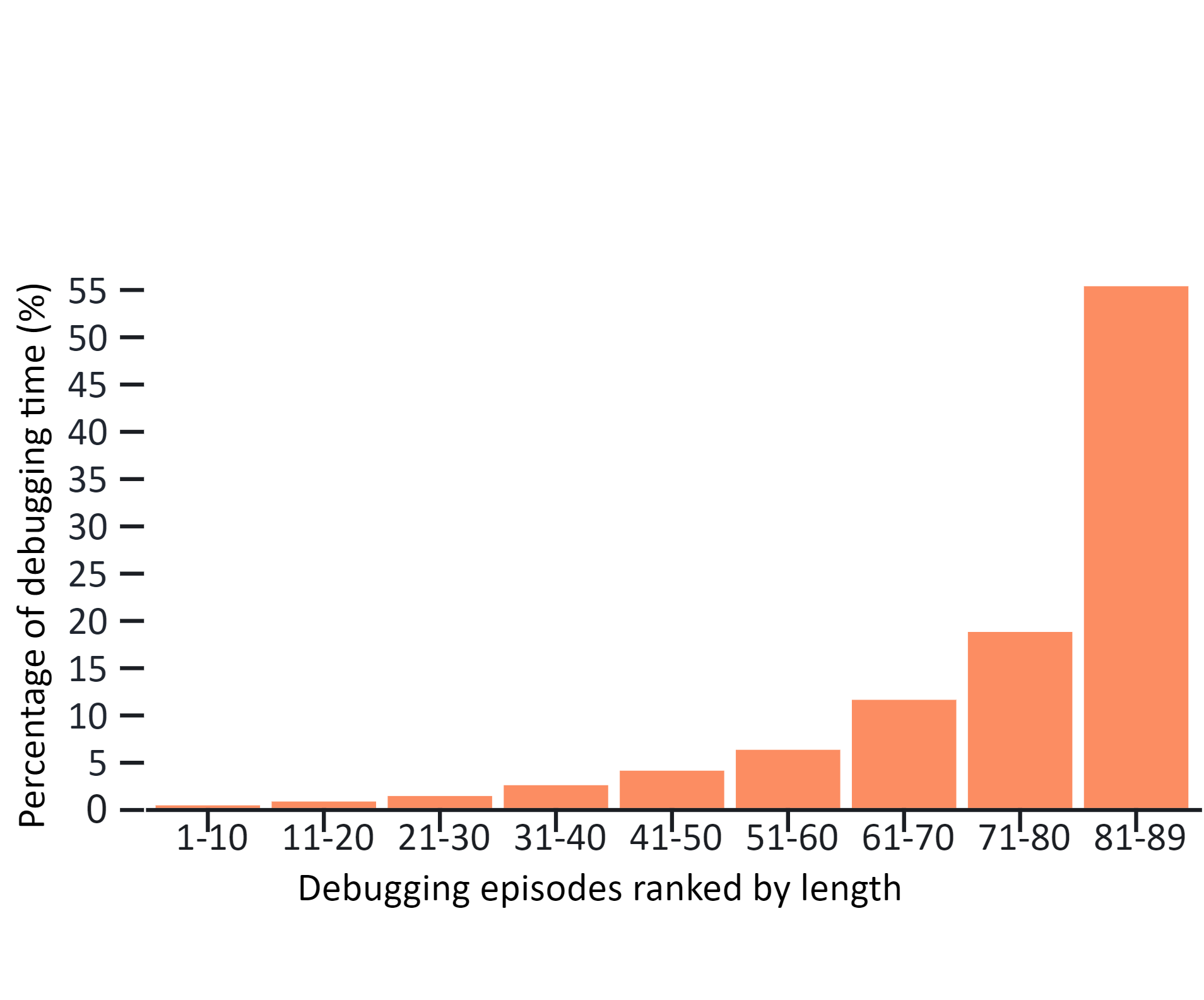}
    \caption{Debugging episodes were sorted from shortest to longest and then grouped in tens. The majority of the total time came from small number of episodes. }
    \label{fig:episodesTime}
\end{figure}
\subsection{RQ1: Frequency and Duration of Debugging Episodes}
Overall, debugging episodes constituted 13\%-95\% (avg = 48\%, sd = $\pm$24\%) of sessions time. This is considerably higher than estimates derived from log analysis of debugger usage (14\%). However, it is closer to developers' self-reported (20\%-60\%) \cite{beller2018dichotomy}. 

We categorize debugging episodes into two categories based on the defect source. The first category contains episodes related to reported defects. Ten debugging episodes started because five developers (D1, D2, D3, D5, and D8) worked towards debugging reported defects in the issue tracker. These episodes consumed most of the session time (79$\pm$16\%). This may not be a surprising observation since developers dedicated the entire session to these reported defects. The second category concerns episodes related to inserted defects. In the remaining 79 debugging episodes, developers (D3, D4, D6-D11) started debugging because they inserted defects while implementing new features. Although developers' goal was to add new features, they spent on average 40\% ($\pm$12\% ) of the session time on debugging.

Developers spent an average of 51 ($\pm$38) minutes debugging each reported defect. In contrast, developers spent 6 ($\pm$8) minutes debugging defects they had just inserted themselves. One may conclude that due to differences in length, debugging episodes concerning reported defects were more problematic than those concerning inserted defects. Although debugging episodes concerning inserted defects were generally short, they were also frequent. While programming, developers constantly debugged new defects, on average after 8 ($\pm$10) minutes of programming. These debugging episodes also were not always short; 20\% lasted for tens of minutes (18$\pm$9). Not all debugging episodes concluded with a successful fix, suggesting that some might be longer if the developers continued. Half of debugging episodes triggered by reported defects did not end with a successful fix, either because more information was needed to reproduce the defect or the developer deferred debugging until later. For new defects created while programming, 14\% did not conclude with a successful fix, either because the developer had higher priority tasks (e.g., shipping the feature even it is not completely correct) or deferred the work until later. Overall, the debugging episodes' distribution was skewed with 80\% of the total episodes time stem from 25 episodes (Figure \ref{fig:episodesTime}).

\begin{table*}[]
\centering
\caption{A summary of the percentage of debugging episode time as well the frequency per debugging episode for each activity. The distribution of occurrence across episode Time shows in which part of a debugging episode the activity occurred the most.}
\label{tab:summeryActivities}
\resizebox{\textwidth}{!}{%
\begin{tabular}{p{4.4cm}ccc}
\hline
&
  \textbf{Browsing A File of Code} &
  \textbf{Editing A File of Code} &
  \textbf{Testing Program} \\\toprule
 \% of Debugging Episode Time
 &
  0\%-50\% (avg = 12\%, sd = ±12\%) &
  0\%-97\%  (avg = 40\%, sd = ±21\%) &
  0\%-100\%  (avg = 40\%, sd = ±21\%) \\ 

Frequency Per Debugging Episode &
   \multicolumn{1}{c}{0-109 (avg = 7, sd = ±14) } &
   \multicolumn{1}{c}{0-67 (avg = 7, sd = ±10)} &
  0-32 (avg = 6, sd = ±6) \\
   Distribution of Occurrence Across Episode Time &
   \multicolumn{1}{c}{{\includegraphics[scale=.04, trim = 5cm 8cm 10cm 4cm]{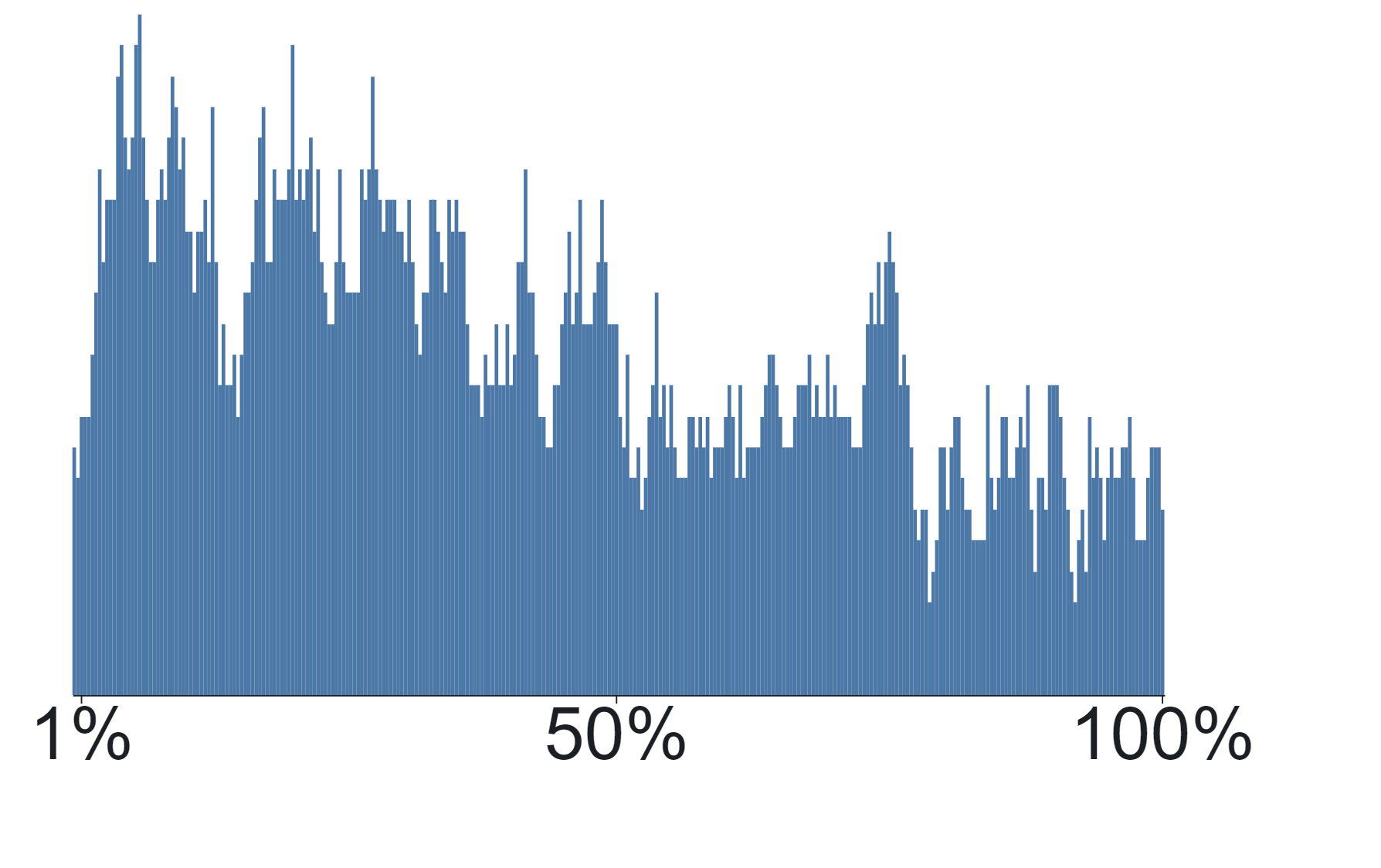}}} &
   \multicolumn{1}{c}{{\includegraphics[scale=.04, trim = 5cm 8cm 10cm 4cm]{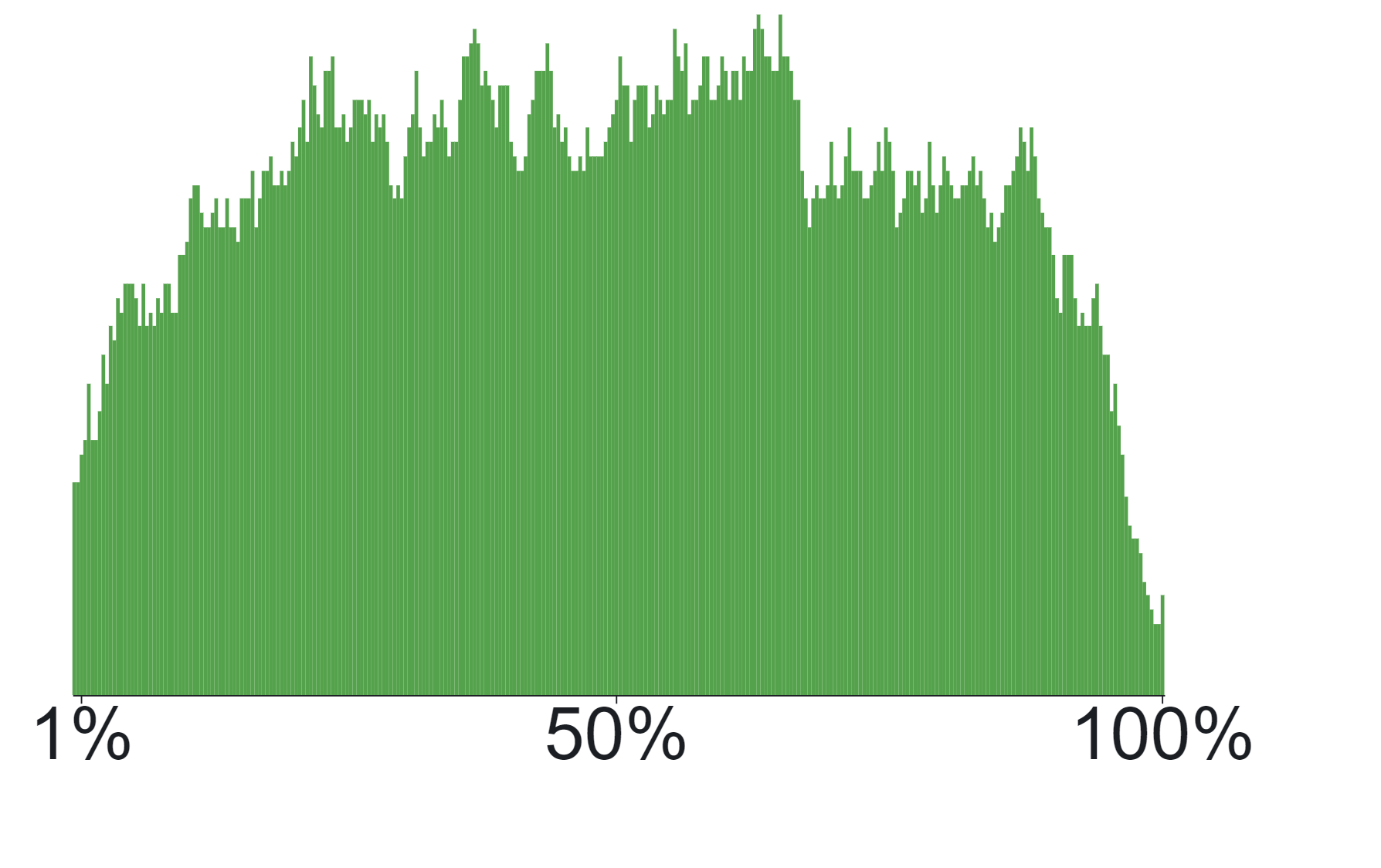}}} &
 {\includegraphics[scale=.04, trim = 5cm 8cm 10cm 4cm]{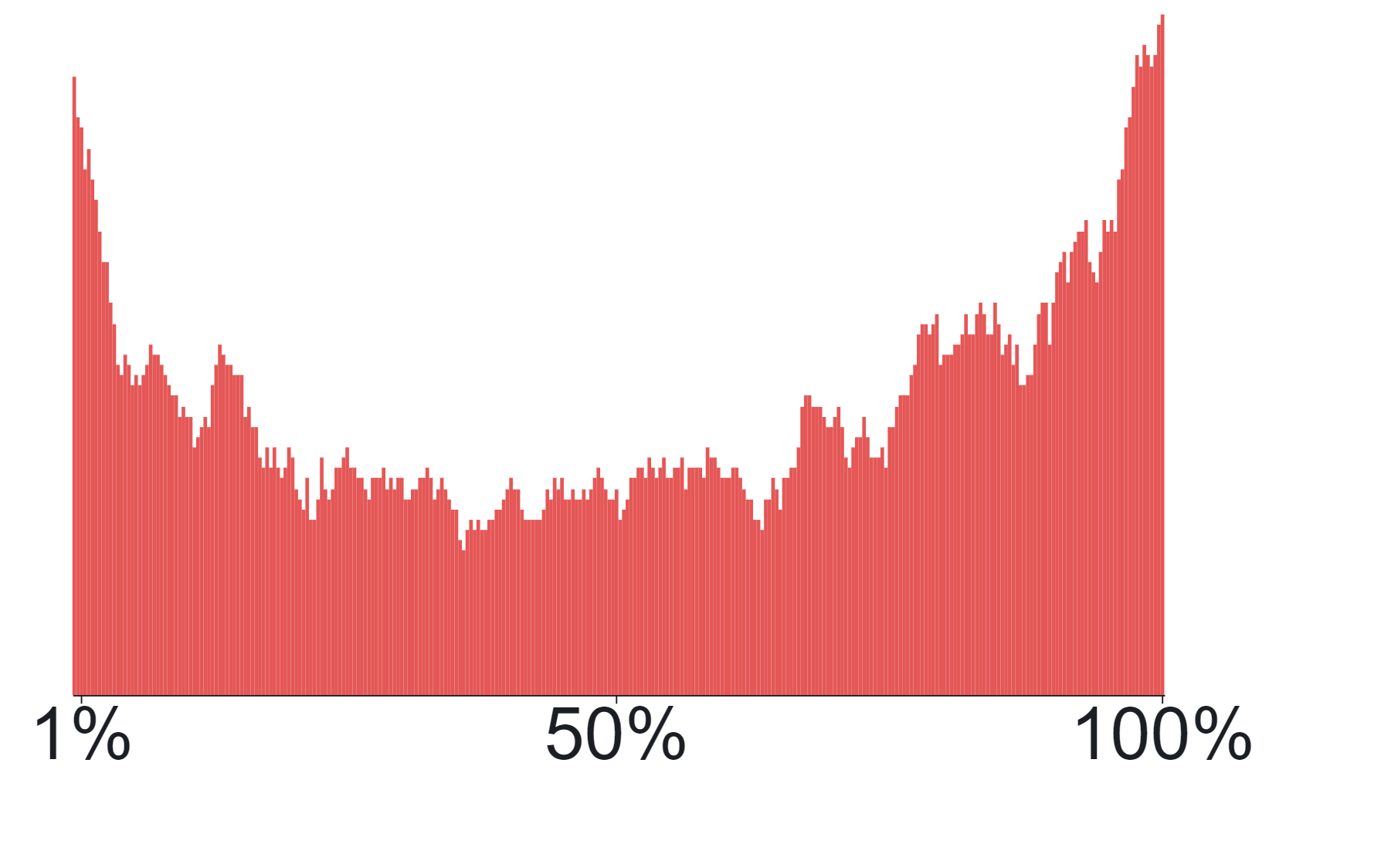}}  \\

\midrule
 
 &
  \textbf{Inspecting Program State} &
  \textbf{Consulting External Resources} &
  \textbf{Other} \\\toprule
 \% of Debugging Episode Time
 &
 0\%-58\%  (avg = 8\%, sd = ±14\%) &
 0\%-59\%  (avg = 3\%, sd = ±10\%) &
 0\%-49\%  (avg = 4\%, sd = ±8\%) \\ 

Frequency Per Debugging Episode  &
  \multicolumn{1}{c}{0-26 (avg = 2, sd = ±14)} &
  \multicolumn{1}{c}{0-16 (avg = 1 ±2)} &
   0-39 (avg = 2, sd = ±5) \\
 Distribution of Occurrence Across Episode Time &
   \multicolumn{1}{c}{{\includegraphics[scale=.04, trim = 5cm 8cm 10cm 4cm]{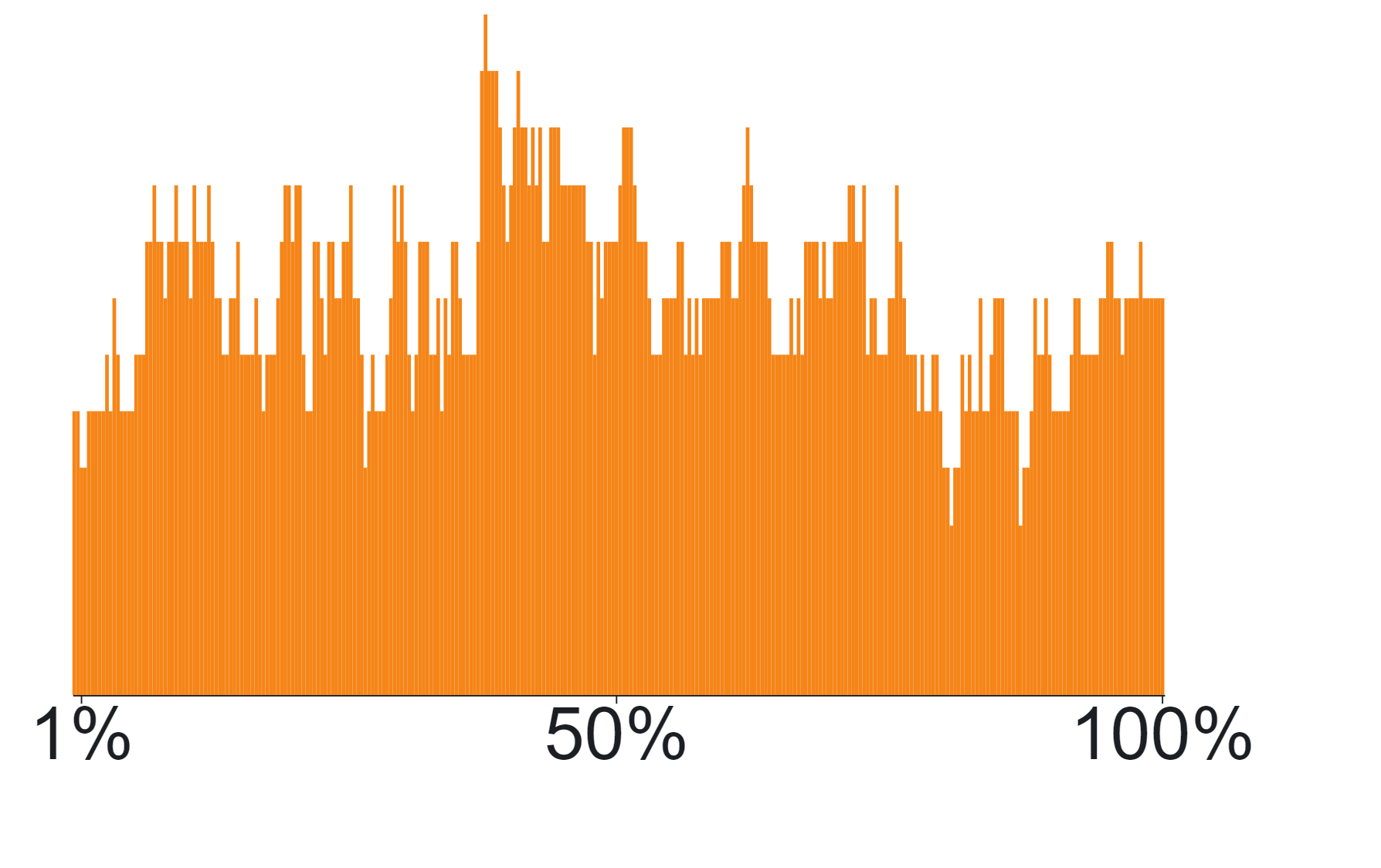}}} &
   \multicolumn{1}{c}{{\includegraphics[scale=.04, trim = 5cm 8cm 10cm 4cm]{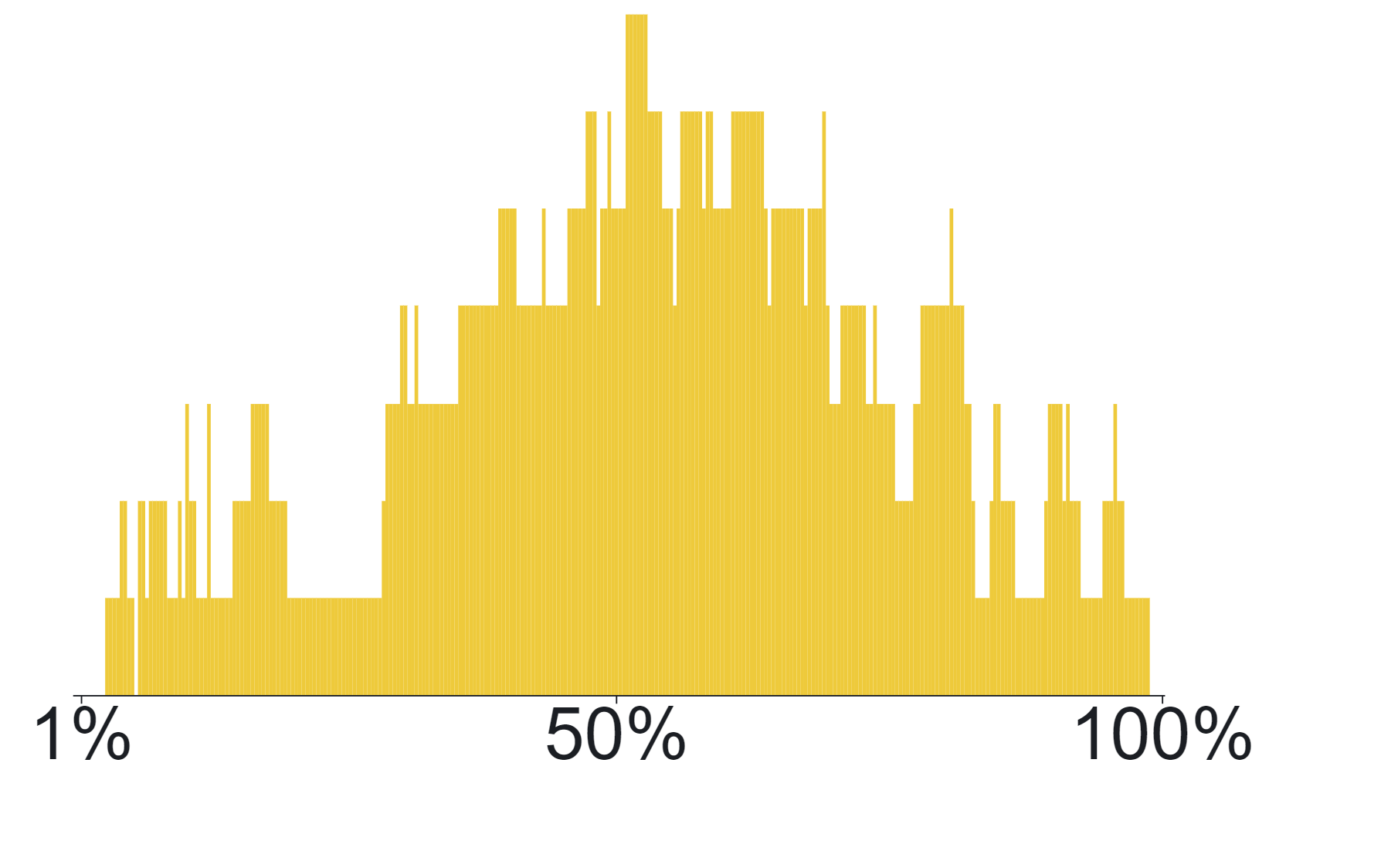}}} &
  {\includegraphics[scale=.035, trim = 5cm 10cm 10cm 4cm]{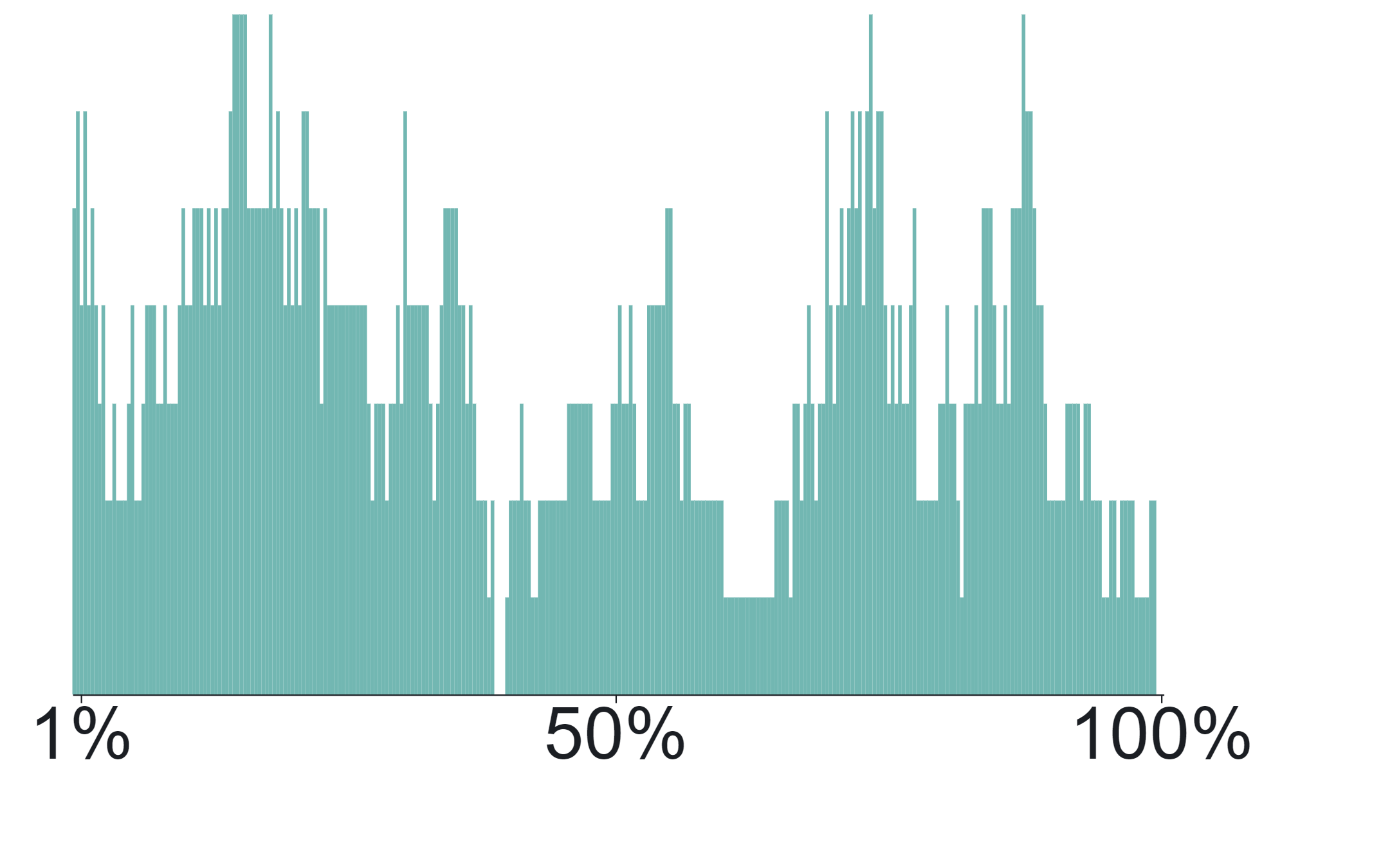}}  \\
    \bottomrule
\end{tabular}%
}
\end{table*}
\begin{table}[]
\centering
\caption{Characteristics of debugging activities.}
\label{tab:my-table}
\begin{tabular}{lll}
\hline
\multicolumn{1}{c}{Activity} &
  \multicolumn{2}{c}{Characteristics} \\ \toprule
Browsing A File of   &
    Duration &
  14s (±21s)/occurrence \\ 
  Code &
  \# of files &
  3($\pm$5)/episode \\
 & 
  File type &
  \begin{tabular}[c]{@{}l@{}}Source code 85\%, \\ Test 7\%, Config 8\%\end{tabular} \\ \hline
Editing A File of &
    Duration &
  33s (±63s)/occurrence \\ Code &
  \# of files &
  2 ($\pm$2)/episode \\
 &
  File type &
  \begin{tabular}[c]{@{}l@{}}Source code 84\% \\ Test 8\%, Config 8\%\end{tabular} \\
 &
  Type of edits &
  \begin{tabular}[c]{@{}l@{}}Functional 94\%\\ Instrumental 6\%\end{tabular} \\ \hline
Testing Program &
    Duration &
  22s (±26s)/occurrence \\ &
  Methods &
  \begin{tabular}[c]{@{}l@{}}Manual 84\%\\ Unit tests 16\%\end{tabular} \\ \hline
Inspecting Program  & Duration &
  53s (±105s)/occurrence \\ State &
  Methods &
  \begin{tabular}[c]{@{}l@{}}Log 70\%\\ Debugger 30\%\end{tabular} \\ \hline
\multirow{2}{*}{\begin{tabular}[c]{@{}l@{}}Consulting External \\ Resources\end{tabular}} &
 Duration &
  35s (±42s)/occurrence \\ &
  Resources &
  \begin{tabular}[c]{@{}l@{}}Doc 83\%, Q\&A 19\%\\ Code example 9\%\end{tabular} \\
 &
  Causes &
  \begin{tabular}[c]{@{}l@{}}Explanation of: \\ Error 8\%, API 92\%\end{tabular} \\ \hline
Other &
 Duration &
  22s (±24s)/occurrence \\ &
  Type of work &
  \begin{tabular}[c]{@{}l@{}}Interaction w/:\\ IDE 74\%, Notes 3\% \\ Issue tracker 15\%\end{tabular}
  \\
  \bottomrule
\end{tabular}
\end{table}


\begin{figure}[ht] 
  \begin{subfigure}[b]{1\linewidth}
           \centering
             \includegraphics[scale=.29, trim = 2cm 1cm 0cm 0cm]{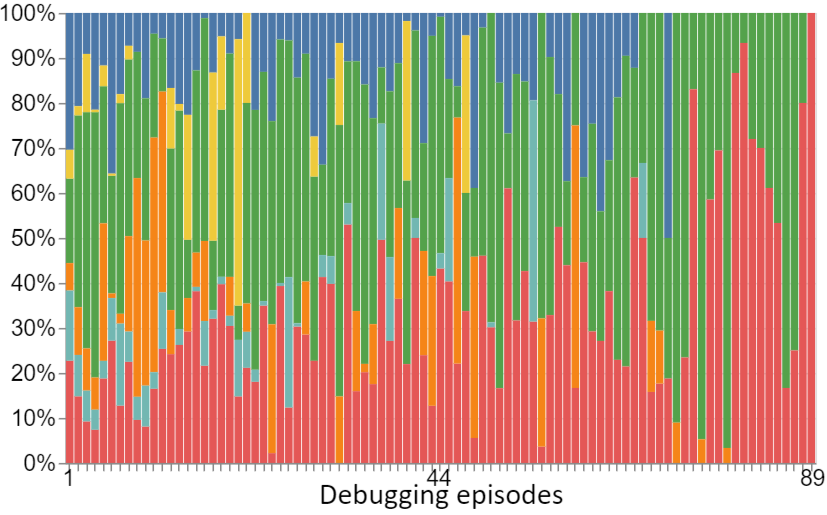}
  \end{subfigure}
    \begin{subfigure}[b]{1\linewidth}
          \centering
         \includegraphics[scale=.21, trim = 3cm 1cm 0cm 0cm]{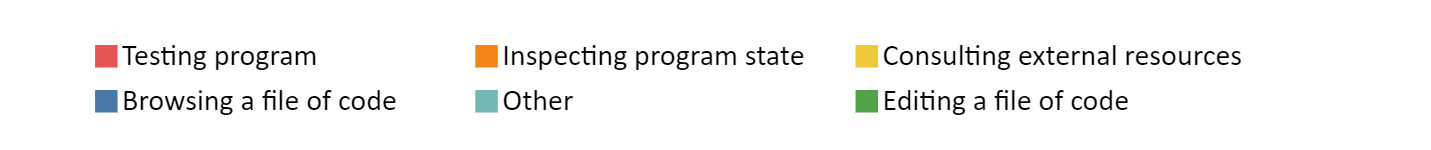}
  \end{subfigure} 
   \caption{The fraction of time spent in each activity per debugging episode.}
\label{fig:episodesActiviries}
\end{figure}
\subsection{RQ2: Activities in Debugging Episodes}

We found that debugging episodes were diverse in their activity: 
there was no single activity that dominated all debugging episodes.
Instead, the time developers spent on each activity varied widely between debugging episodes (Figure \ref{fig:episodesActiviries}). 
Table \ref{tab:summeryActivities} summarizes the occupancy and occurrence of each activity. Table \ref{tab:summeryActivities} summarizes key characteristics of each activity.

\textit{Browsing a file of code} activity occurred, on average, 7 ($\pm$14) times per episode, occupying 12\% ($\pm$12\%) of the episode time. 
Developers browsed 3 ($\pm$5) distinct files per episode (max=32 files). 67\% ($\pm37$\%) of the files that developers browsed during a debugging episode were later edited during the same episode. After finishing browsing a file of code, developers were most likely to visit another file to browse (40\%) or edit (24\%). 

\textit{Editing a file of code} activity occurred an average of 7 ($\pm$10) times during debugging episodes, for an average 40\% ($\pm$21\%) of episode time. 
Developers edited 2 ($\pm$2) files of code for each episode (max=13). When developers edited only a single file, they usually did not complete their edit in one time segment. Developers instead switched back and forth an average of three times between editing that file and other debugging activities.
Developers were most likely to test their edit (58\%) or inspect program state (12\%).

Developers engaged in an average of 6 ($\pm$6) \textit{testing program} activities per debugging episode. They spent 34\% ($\pm$22\%) of episode time on this activity. Developers most often ran and observed the program output manually (84\%) rather than through automated tests (16\%).
Developers were likely to next either edit a file of code (56\%) or browse a file of code (33\%).

\textit{Inspecting program state} occurred, on average, twice ($\pm$14) per debugging episode, for 8\% ($\pm$14\%) of debugging episode time. 
To inspect the program state, developers used a combination of log statements (70\%) and breakpoints (30\%). Developers spent 53 ($\pm$105) seconds each time they inspected program state, the longest instance duration of any activity. Developers were likely to edit (65\%) or browse a file of code (24\%) after inspecting the program state.

Developers engaged in an average one ($\pm$2) times in \textit{Consulting external resources} activities per debugging episode. This activity was the least common, occurring in only 21\% of debugging episodes. However, 91\% of developers consulted external resources at least once. 
When consulting external resources, developers primarily searched for an explanation of an API (92\%) rather than an explanation of defect behavior (e.g., an error message)(8\%). The most common information source was API documentation (83\%). Developers also sought existing code examples (9\%) and relevant posts in Q\&A communities(19\%). After consulting external resources, developers often edited (39\%) or browsed (35\%) a file of code.

\textit{Other} activity occupied on average 4\% ($\pm$ 8\%) of the debugging episode time and 
In this activity, the mechanics of interacting with the development environment to search for keywords inside files, navigate between files and folders constitute 74\% of what developers did. Developers also browsed issue trackers (15\%) and took note (3\%) as part of the other activity.

We also examined the order of debugging activities within debugging episodes. We found that there was not a single point within debugging episodes when activities occurred. For example, developers browsed  files of code anywhere from the start to the end of debugging episodes. However, there were \textit{peak} occurrences for most debugging activities when they were most common~(Table \ref{tab:summeryActivities}). Testing was most common at the beginning and end of debugging. One explanation is that developers test when beginning to debug to understand the defective behavior and at the end of debugging to confirm that it produces the output they code. Browsing a file of code was more common during the first half of debugging episodes. This might be because developers needed to collect relevant code or localize the defect before engaging in different activities. Editing source code was widely distributed across the debugging episode, peaking in the middle of episodes. Consulting external resources exhibited a strong peak in the middle of debugging. This may correspond to the point when developers first try to fix a defect but get stuck and decide to seek additional information elsewhere. Another explanation is that developers may start implementing a fix and later consult help to understand how to implement the fix. Inspecting program state and other activities were widely spread across debugging episodes.

\begin{figure*}[ht] 
  \begin{subfigure}[b]{0.5\linewidth}
           \centering
             \includegraphics[scale=.5,  trim= 0cm 1.5cm 0cm .5cm,clip]{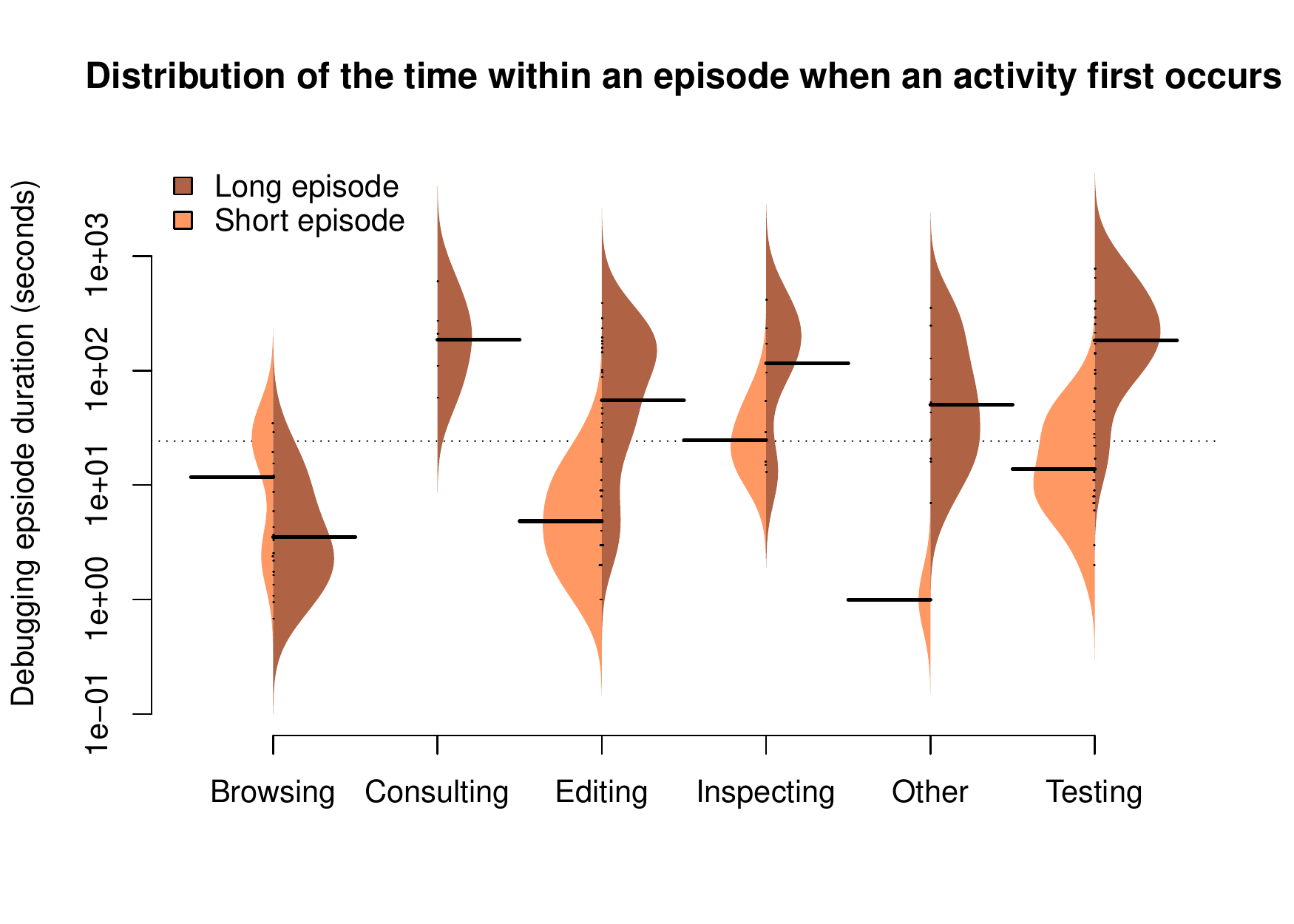}
  \end{subfigure}%
  \begin{subfigure}[b]{0.5\linewidth}
          \centering
            \includegraphics[scale=.5,  trim= 0cm 1.5cm 1.5cm .5cm,clip]{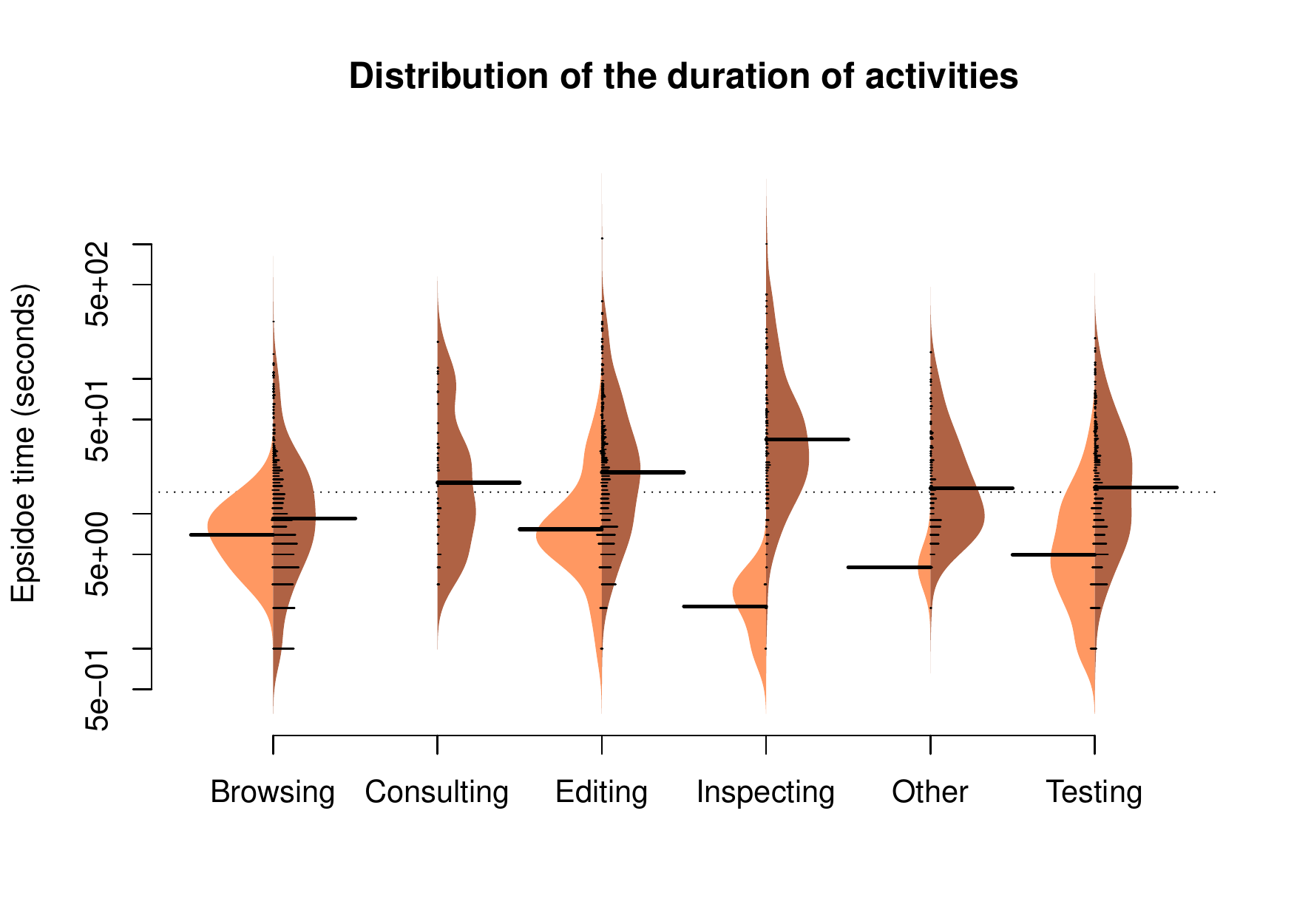}
  \end{subfigure}
           \caption{The distribution of the time within an episode when an activity first occurs, in log scale (left). The distribution of the duration of activities, in log scale (right). The thick black lines indicate the mean.}
    \label{fig:longVsShort} 
\end{figure*}


\subsection{RQ3: Activities in Long and Short  Debugging Episodes}

To investigate the differences between long and short debugging episodes, we calculated the 75th and 25th percentile of debugging episode durations. We marked any debugging episode that lasted for 12.3 minutes or more (\textgreater= 75th percentile) as \textit{long} and any episodes with a duration less than a minute (\textless= 25th percentile) as \textit{short}. This resulted in 23 episodes in each group. The 23 short debugging episodes (31$\pm$17 seconds) were from eight developers (D3-D9, D11) and constituted only 1\% of overall debugging time. In contrast, the 23 long debugging episodes (32$\pm$23 minutes) were from ten developers (D1-D5, D7-D11) and constituted 80\% of all debugging time. The long debugging episode threshold that we defined aligns with what professional developers report as the 15-minute threshold at which they begin to perceive debugging as difficult \cite{BohmeFSE-DebuggingHypotheses}. As long debugging episodes were longer in duration, they involved many times more activities (62$\pm56$) than short episodes (4$\pm2$). Therefore, we focus on examining the fraction of each episode's time spent in each activity. 

\begin{figure}[ht] 
  \begin{subfigure}[b]{1\linewidth}
    \centering
    \includegraphics[scale=.13,trim=9cm 4cm 10cm 3cm,clip]{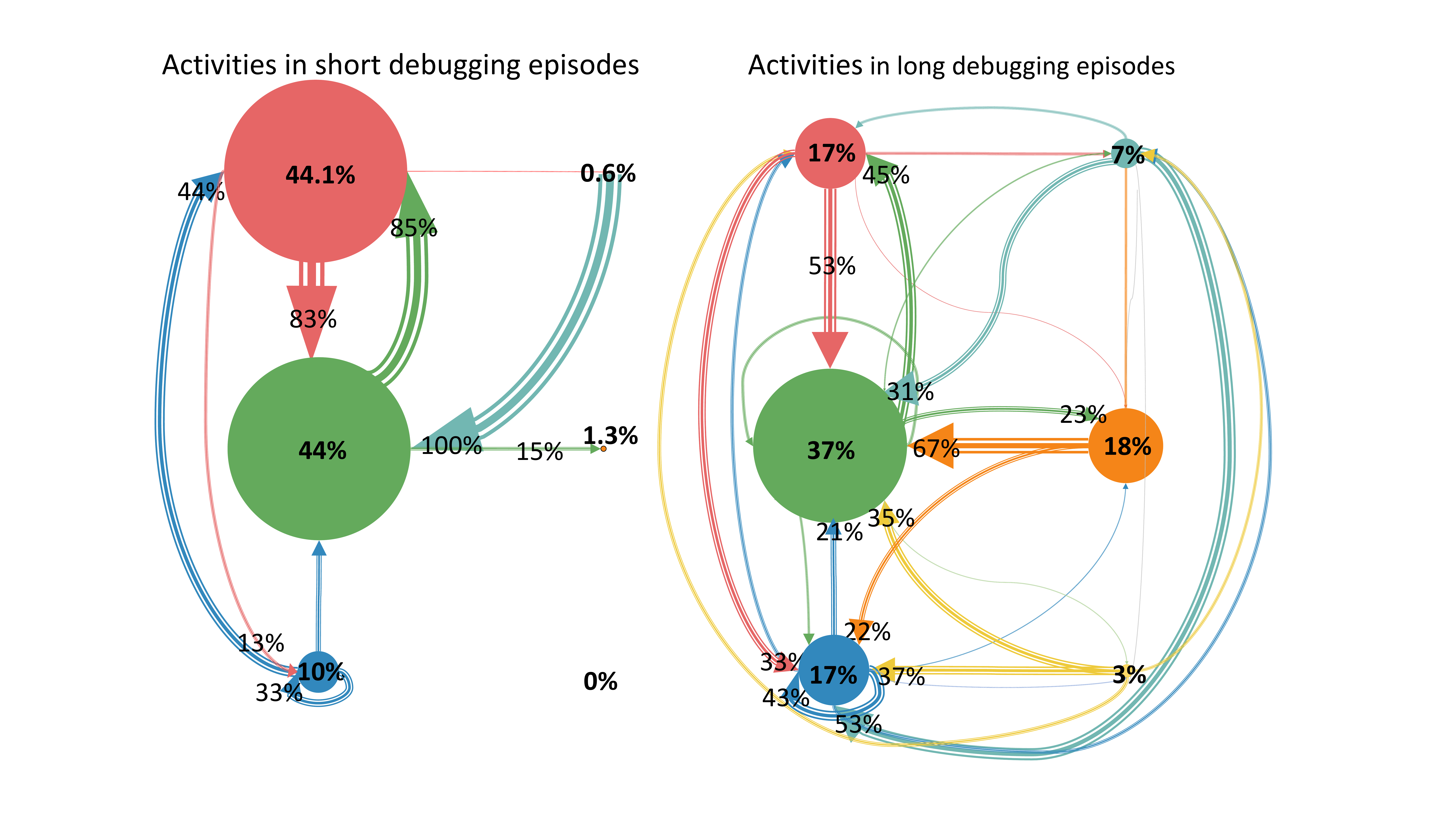}
  \end{subfigure}
   \begin{subfigure}[b]{1\linewidth}
          \centering
         \includegraphics[scale=.21, trim = 3cm 1cm 0cm 0cm]{figures/legends.png}
  \end{subfigure} 
 \caption{Developer activity in short and long debugging episodes. The circle widths signify the fraction of time spent on each activity. The lines (with the color of the corresponding activity) describe the fraction of transitions to the next activity (indicated by line width). Transition percentages are labeled for the two most common transitions.}  
 \label{fig:shortVsLongActivity}
\end{figure}


We observed two key differences distinguish long debugging episodes from short episodes. First, long debugging episodes involved a diverse set of activities, with developers switching between different type of activities (Figure \ref{fig:shortVsLongActivity}). Second, developers spend more time on each activity instance in long debugging episodes. They were also slower in making their first edit to the source code (Figure \ref{fig:longVsShort}).  

Developers engaged in a more diverse set of activities in long debugging episodes. Short debugging episodes mostly focused on editing and testing, occupying 88\% of their time. Developers edited an average of 1($\pm$.2) files very early in the episode (after 6 ($\pm$8) seconds) and spent only 11 ($\pm$11) seconds on each edit to a file. Developers tested their program an average of 2 ($\pm$1) times for an average of 8 ($\pm$8) seconds each time. Editing a file of code and testing program activities constituted only 54\% of the total debugging time on the long debugging episodes. However, developers edited more files than in the short debugging episodes. On average, developers edited three files ($\pm$3, max = 13) and spent on average 41 (75$\pm$) seconds each time they open a file to edit it. Developers were also slower to first edit in long debugging episodes. Developers spent 3.2 ($\pm$6, max = 30) minutes before introducing any changes to the source code. We found that developers tested their program more in longer debugging episodes. On average, developers tested their program 12 ($\pm$8) times for an average of 26 ($\pm$31) seconds each time. 

Developers spent more time (17\%) browsing a file of code in long debugging episodes (Figure \ref{fig:shortVsLongActivity}). This may be related to two factors. First, developers in long debugging episodes browsed more files (7$\pm$7, max = 32 ) than the short debugging episodes (0.2$\pm$.8, max = 3). Second, developers spent more time each time they browsed a file. In long debugging episodes, developers spent an average of 17 ($\pm$24) seconds each time they browsed a file compared to an average of 8 ($\pm$4) seconds in short debugging episodes.   

The amount of time spent inspecting program state varied the most in long debugging episodes, increasing from 1.3\% to 18\% in long debugging episodes. In short debugging episodes, inspecting program activity was not followed by another activity (i.e., there was no arrow from inspecting program activity to any other activity). However, in long debugging episodes, editing a file of code and inspecting program activities had the second most switches between activities. Developers engaged the most extended time per instance in inspecting program state activity than other activity (1.2$\pm$2 mins, max=3 mins) in long debugging episodes.

Consulting external resources and other activities were very rare in short debugging episodes, with only one instance in our dataset. However, consulting external resources and other activities were common in long debugging episodes (48\% and 83\%, respectively) and constituted 3\% and 7\% of the time.

Overall, we found that the differences between short and long debugging episodes were due to developers switching between different activities and spending more time each time they do. No activity emerged as a singular bottleneck that accounted for the majority of the time in long debugging episodes. 





\begin{figure*}[ht] 
 \begin{subfigure}[b]{0.5\linewidth}
           \centering
             \includegraphics[scale=.5,  trim= 0cm 1.5cm 0cm .5cm,clip]{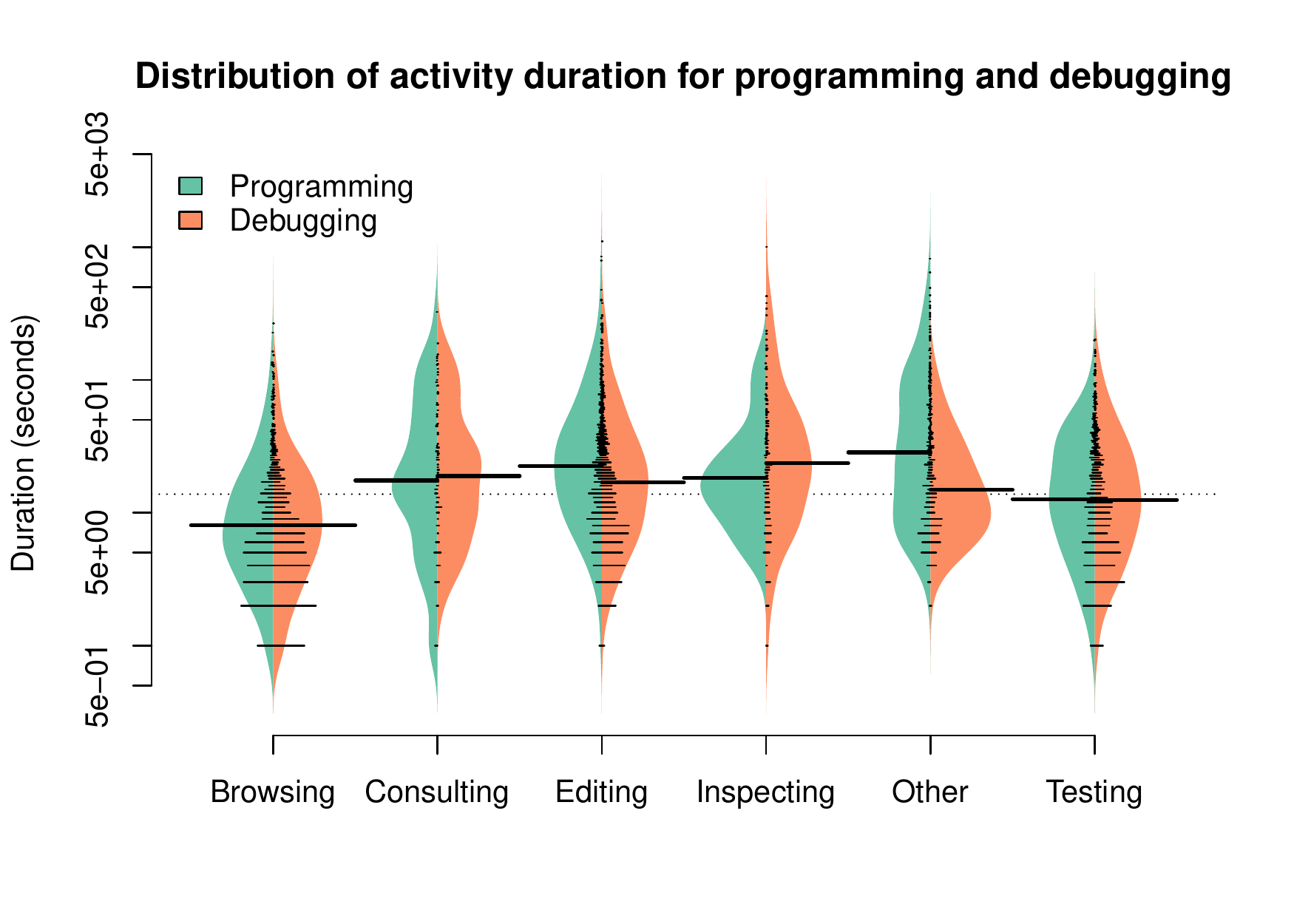}
  \end{subfigure}%
  \begin{subfigure}[b]{0.5\linewidth}
          \centering
            \includegraphics[scale=.43,  trim= 0cm .5cm 0cm 0cm,clip]{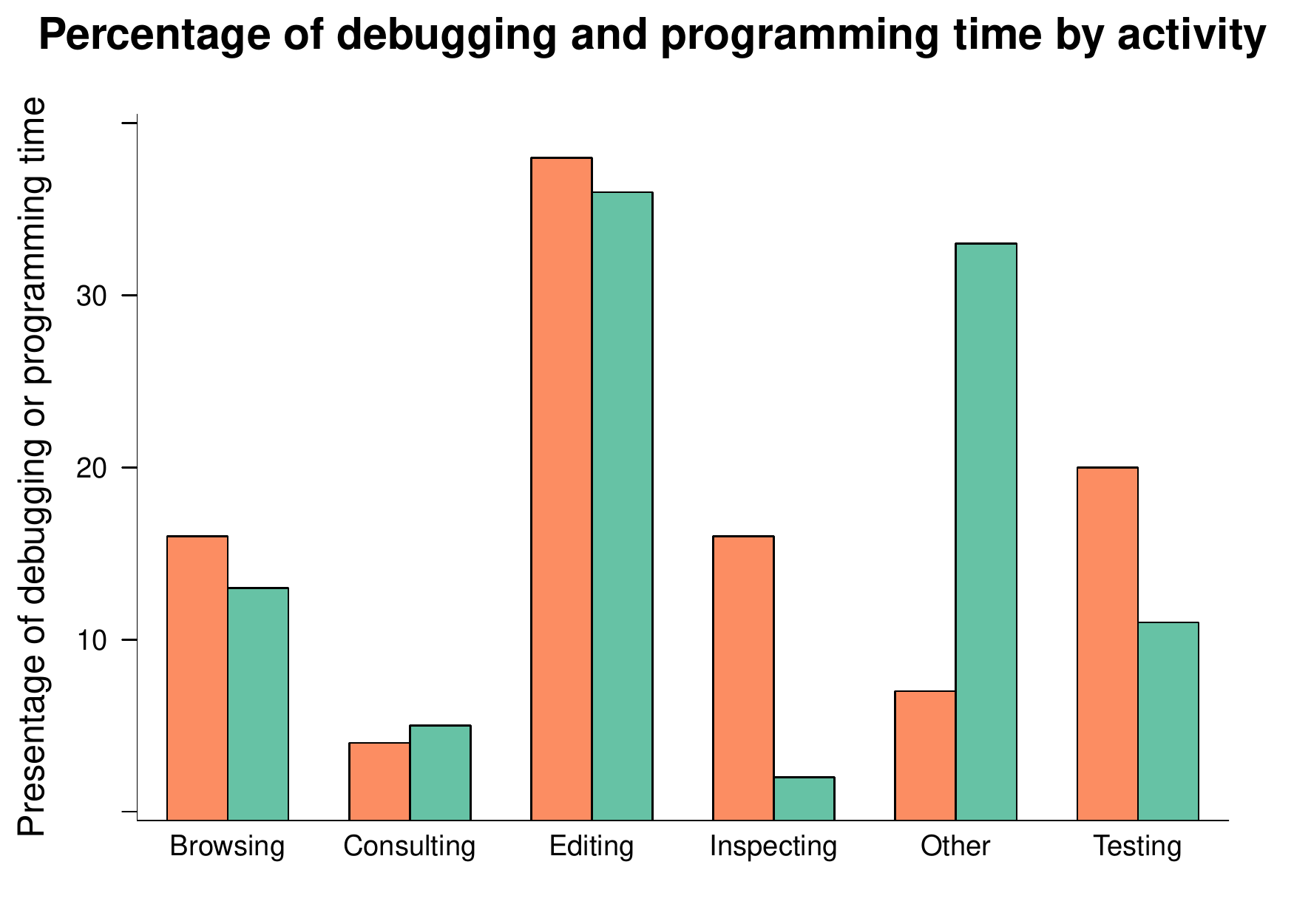}
  \end{subfigure}

           \caption{The distribution of activity duration for programming and debugging work, in log scale (left). The thick black line indicates the mean. The percentage of debugging and programming time by activity (right). }
    \label{fig:programVsDebug} 
\end{figure*}

\subsection{RQ4: Activities in Debugging Episodes and Programming Work}
To investigate how debugging episodes vary from programming work, we compared the 1407 activities in the 13 hours of programming to the 2137  activities in the 15 hours of debugging. We found debugging and programming were broadly similar in the fraction of time developers spent browsing and editing code files and consulting external resources. There were more pronounced differences in the time developers spent testing and inspecting code. Figure \ref{fig:programVsDebug} plots the distribution the overall differences and similarities between programming and debugging activities. 

Three activities were most similar between debugging and programming work. Developers spent 16\% of their debugging time browsing a file of code, compared to 13\% of their programming time. Developers spent 14($\pm$21) seconds browsing a file of code while debugging compared to 15($\pm$22) seconds while programming. Editing a file of code activity constituted 38\% of debugging time and 36\% of the programming time. Developers spent 33($\pm$63) seconds and 44($\pm$78) seconds each time they edited a file of code in debugging and programming work, respectively. Developers spent similar fractions of their debugging (4\%) and programming (5\%) work consulting external resources, with similar durations for debugging (35$\pm$42 seconds) and programming (39$\pm$52 seconds).

Programming and debugging work were less similar in the time spent testing and inspecting program state, mostly as developers did these more often when debugging. When debugging, developers switched to testing twice as often and inspecting the program state six times more often than when programming. Developers were more likely to test their changes (58\%) or inspect the program state (12\%) after finishing editing a file of code while debugging. However, developers switching behavior changed in programming work. After editing a file of code,  developers often sought to edit another file of code (25\%) or test their changes (36\%). Besides the differences in switching behavior, the time developers spent on each instance of inspecting program activity was longer in debugging work (53s$\pm$105s) than in programming (28s$\pm$32s). However, the duration of testing activities in debugging (22s$\pm$26s) and programming (21s$\pm$22s) was similar.

In programming and debugging, developers spent 61\% and 74\% of the other activity, respectively, while interacting with their development environment, navigating file systems, installing libraries, setting up their IDE, and opening up other tools and software. Browsing the issue tracker was the second most common type of other activity. In programming work, developers used the issue tracker three times more often than in debugging. The differences suggest that debugging is a more code-focus activity, while programming work involves much work relating to visiting the issue tracker, writing notes, and setting up the development environment.

\section{Limitations and threats to validity}
Our study has several important limitations.
\textbf{Construct Validity.} Defining exactly when debugging episodes and activities start and end is challenging and potentially susceptible to human error. To minimize the risk incorrect codes, we collected past definitions of debugging \cite{johnson1982software,ko2011state,Parnin2011AreAutomatedDebugging} and used these to create an initial definitions. As we built the initial coding scheme, we refined the definitions until two authors were able to independently and consistently annotate the start and end of video segments within one to two seconds. 

\textbf{Internal Validity.} It has long been known that developers are frequently interrupted in programming work \cite{meyer2014software, abad2018task}, increasing the time needed to finish tasks. To ensure our measures of debugging activities did not include any irrelevant worked caused by interruptions, we coded any interruption that lasted more than five seconds and excluded from the debugging and programming work. We defined the five second threshold after we observed that interruptions that lasted less than that did not cause the developers to pause and switch context.  

\textbf{External Validity}. Researchers have found that live-streamed programming is a source of data that show developers working on open source project. However, these videos may also contain non-trivial of interruptions caused by other developers watching, which may not represent how developers typically work. To mitigate this issue, we used a strict inclusion criteria where a video has to show a significant work and less interactions with other developers watching. We have also other criteria that ensure that these videos showed developers working in nontrivial projects that have been used in production. Another potential threat to external validity is how representatives the sample of developers we observed. We sampled developers with a high range expertise levels, ranging from seven to 31 years of committing to open source code. However, this measure is only a proxy of developers' overall years of experience, which is itself a proxy of expertise.




\section{Observe-Dev.online platform}
\label{sec:Observe.dev}
\begin{figure}
    \centering
    \includegraphics[scale=.33, trim= 0cm 1cm 10cm 0cm]{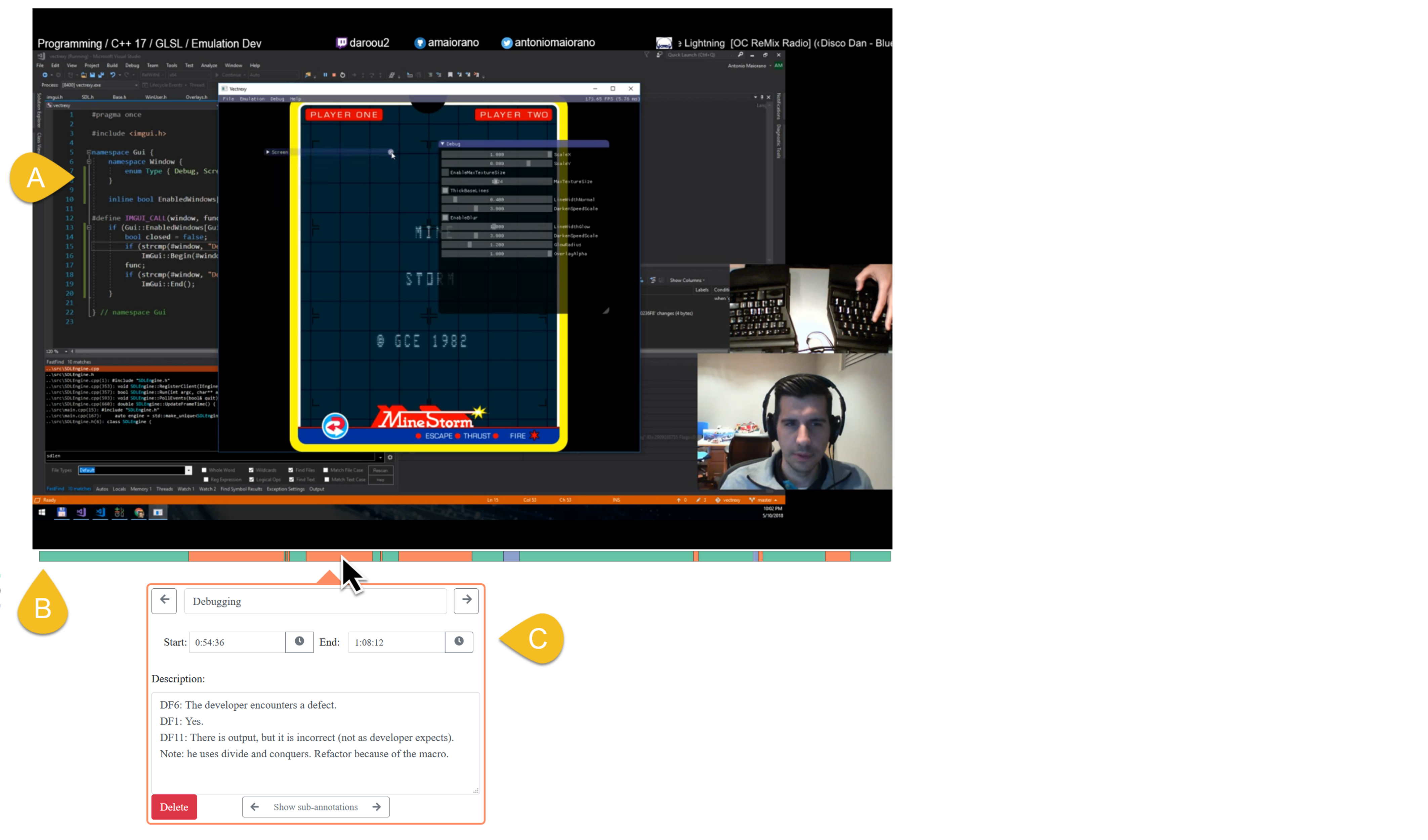}
  \caption{Observe-dev.online supports collaborative qualitative analysis of lives-streamed programming videos. Programming videos (A) are shown with an interactive timeline view (B) of video segment annotations supporting navigating between segments. New codes can be edited (C) or created based on the current position in the video. }
    \label{fig:observe.dev}
\end{figure}

Live-streamed programming offers an important opportunity for researchers to observe professional developers in a natural setting. Conducting studies with similar settings would require researchers to conduct field studies and record developers' screens and voices during the work. Live-streamed programming is an alternative that requires no such effort with public access to both the source code and recording. As we used this rich source of information to explore debugging, we believe that researchers can use these videos to explore further debugging or observe developers exploring other research areas related to developers' programming behavior. For instance, researchers may investigate the use of online resources during debugging and programming work, the challenges developers face working in specific programming languages, or how developers make design decisions. We built a platform that supports software engineering researchers in organizing videos, collaboratively analyzing and annotating videos, and sharing datasets both publicly and privately. Figure \ref{fig:observe.dev} depicts the platform interface for annotating a live-streamed programming video. 

Observe-Dev.online offers four key features for supporting the use of live-streamed programming videos in software engineering research.
First, the platform offers \textit{a dataset of programming sessions}. Identifying live-streamed programming videos can be time-consuming, particularly to identify videos with specific characteristics (e.g., working on a data analysis script in Python). Therefore, Observe-Dev.online includes a default dataset of more than 100 hours of programming sessions that are publicly available for use\footnote{https://bit.ly/3qWdMVA}. Each session is labeled with metadata, describing the programming languages, projects, and development environments used. Researchers can use this to filter sessions to match inclusion criteria. Second, Observe-Dev.online offers the \textit{ability to annotate video segments}. Researchers can create new codes and annotate specific video segments with these codes. Segments may vary in duration from one second to the entirety of the video time. After applying codes to segments, the tool offers a mini-timeline visualization of the codes, enabling them to see where codes are located at a glance and quickly navigate to specific code locations. For example, Figure \ref{fig:observe.dev} shows three codes for programming, debugging, and irrelevant episodes, each shown in a unique color. Third, researchers can \textit{share annotated datasets}. Observe-dev.online  is a web-based platform, enabling datasets to be publicly or privately shared for viewing or editing through a URL and optional authentication. Finally, the platform supports \textit{exporting annotations} to a standard JSON format that can be imported into other tools for further analysis.

\section{Discussion}



Our observation of debugging episodes offered insight into the nature of debugging in a naturalistic context. Developers spend about half of their programming time debugging, suggesting that debugging clearly remains a core and time-consuming part of programming. Moreover, we found that debugging episodes are surprisingly frequent, where developers debug after every eight minutes of programming work. Analyzing the activities that occurred within debugging episodes revealed that episodes were diverse in activities. Developers may browse files of code more in specific episodes and consulate external resources more in others. Also, activities in longer debugging episodes were more diverse than the shortest debugging episodes. Developers were mostly testing and editing their code in the shortest debugging episodes, while developers in longer debugging episodes had to browse and edit many files, inspect program state, and consulate external resources. Another surprising result emerged after investigating differences between debugging and programming activities: there were similarities between how developers browsed and navigated code in debugging episodes and programming work. There were also differences in developers inspecting and testing program behavior.      

We found that developers' behavior in debugging is surprisingly varied. Further research is needed to understand the reasons for these differences. Investigating these differences in developer behavior might lead to new theories explaining what causes developers to spend more of their time browsing files of code, consulting external resources, or testing while debugging. 

The surprising similarities between debugging and programming behavior merit further investigation. Traditional debugging studies view debugging mostly as a fault localization activity \cite{wong2016survey}. However, our results suggest that developers browse and edit files when both editing and debugging. We also found the differences stem from how they test and inspect the program. Future studies investigating debugging and programming behavior may help understand what debugging is and how it differs from regular programming work.

Our results indicate that developers engaged in a variate of activities in longer debugging episodes. Debugging tools should aim to reduce the potential overhead of switching between many activities. We propose two types of debugging tools.  The first type of tool is \textit{a specialized debugging environment} in which the environment helps developers easily switch between different types of activities. One example of such an environment is the live programming environment. In a live programming environment, developers do not need to switch to another window or program to edit, inspect, or test their program while debugging. The program is continuously evaluated while the developer changes the source code, revealing the program state and outputting the final input for each source code change.  The second type of tool is \textit{an aggregator and summative debugging tools}. Instead of developers switching between many activities, these tools predicate the files developer may need, the resources that developers may consulate, and the program state developers may inspect and present them for developers. These tools differ from existing tools that capture and maintain task context \cite{kersten2005mylar} as they aim to predict and aggregates relevant information within the IDE and on the internet.

\bibliographystyle{IEEEtran}
\bibliography{ref}
\end{document}